\RequirePackage[2020-02-02]{latexrelease}
\documentclass[prb,twocolumn,showpacs,amsmath,amssymb]{revtex4} 
\usepackage{graphicx}
\usepackage{tabularx}
\usepackage{dcolumn}
\usepackage{bm} 
\usepackage{color} 
\usepackage{amsmath}

\begin{document}

\title{Direct ab initio calculation of magnons in altermagnets: 
	   method, spin-space symmetry aspects, and application to MnTe}

\author{L. M. Sandratskii$^{1,2}$\footnote{lsandr3591@gmail.com}, K. Carva$^1$, V. M. Silkin$^{2,3,4}$}
\affiliation{$^1$Faculty of Mathematics and Physics, Charles University, 12116 Prague,
Czech Republic\\
$^2$Donostia International Physics Center (DIPC), Paseo de Manuel Lardizabal 4, E-20018 San Sebasti\'an, Spain\\
$^3$%Departamento de Física de Materiales, Facultad de Ciencias Quimicas, UPV/EHU, 20080 San Sebastian, Spain\\
Departamento de Pol\'{\i}meros y Materiales Avanzados: F\'{\i}sica,
Qu\'{\i}mica y Tecnolog\'{\i}a, Facultad de Ciencias
Qu\'{\i}micas, Universidad del Pa\'{\i}s Vasco (UPV-EHU), Apdo. 1072,
E-20080 San Sebasti\'an, Spain\\
$^4$IKERBASQUE, Basque Foundation for Science, 48011 Bilbao, Spain\\
}

\begin{abstract}
We suggest the \textcolor{black}{first} method for direct ab initio calculation of 
adiabatic magnons in complex collinear
magnets. The method is based on the density-functional-theory (DFT) calculation under 
two different constraints:
one constraint governs the change of the magnetization with respect to the ground
state, and the other is the symmetry constraint responsible for the  value of the 
magnon wave vector. 
\textcolor{black}{
The advantages of the suggested method with respect to the usual approach of mapping of the electron
system on the Heisenberg Hamiltonian of interacting magnetic moments are discussed.
}
The performance of the method is demonstrated by 
the application to an altermagnet MnTe. 
\textcolor{black}{
The altermagnetism introduced as a concept in 2022 is at present an area of highly intensive research.
The characteristic feature of altermagnets is the spin splitting of the electron states in reciprocal 
{\bf k} space. Among the discovered properties of the altermagnets is the chirality splitting of the magnons
in wave vector {\bf q} space. 
We suggest an appoach to the study of the symmetry aspects of magnon chirality splitting.
We show that  both the
chirality splitting of the magnons and the altermagnetic spin splitting of the electron states,  
though very different in their physical nature, 
have identical patterns in the corresponding wave vector spaces.
Since the altermagnetism of MnTe  is the consequence of the 
presence of the Te atoms, an adequate attention is devoted to
the symmetry analysis and calculation results for the Te moments
induced in the magnon states.
In the calculations each magnon is characterized by its own electron band structure. We investigate 
the transformation of the electron structure in the 
transition of the material from the collinear ground state to noncollinear magnon states.
We show the connection between the properties of magnon band structures and the 
chirality properties of magnons.
In the investigation of the chirality splitting
as well as in both the formulation and the application of our method 
an important role play the aspects 
of generalized symmetry based on the application of the spin-space groups. 
The symmetry framework connects in one coherent picture different 
parts of the consideration:
(i) the generalized translational symmetry of the magnons as a crucial condition for their efficient
ab-initio calculation, 
(ii) altermagnetic spin-splitting of the electron states in the ground state, and
(iii) chirality splitting and band structures of the magnon excitations. 
}
\end{abstract}
 
\maketitle
 
\section{Introduction} 
Magnons as low energy excitations of the magnetic systems play a central role in the the magnetic thermodynamics. 
The emerging field of magnonics enhanced further the importance of the magnons \cite{Flebus2024,Chumak2022,Barman2021,Qin2015}.
The first-principles study of the magnons is one of the 
prominent tasks of the theoretical approaches to magnetic systems based on the density functional theory (DFT). 

In the adiabatic picture focusing on the dynamics of atomic magnetic moments, the magnons
are presented by the configurations of the moments 
deviating from the magnetization axis\cite{Kittel_introduction,Halilov1998a,Niu99}. 
\begin{figure}[t]
\includegraphics*[width=5.8cm]{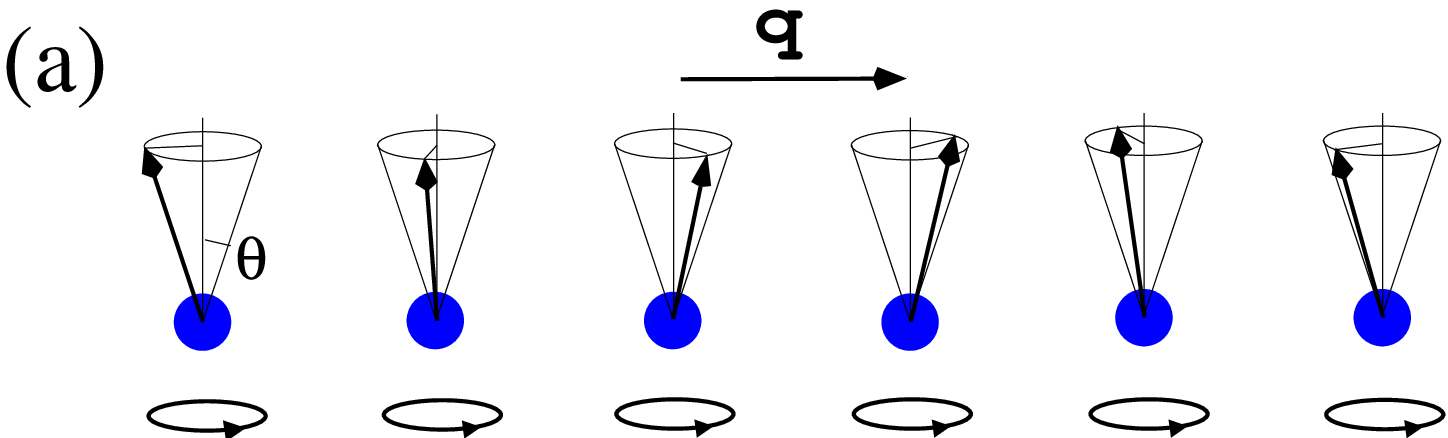} \vspace{0.6cm}\\
\includegraphics*[width=6.0cm]{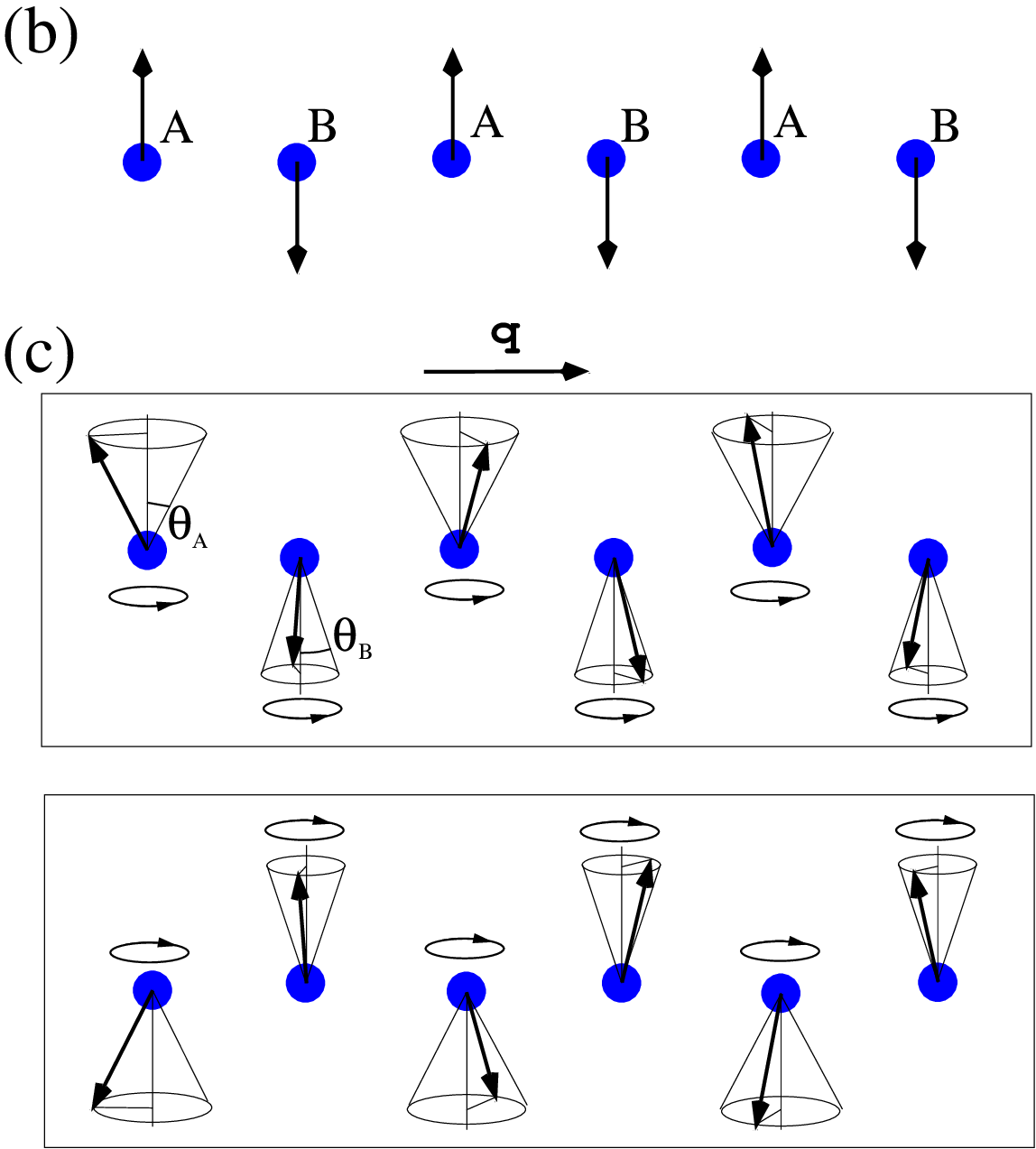} %\vspace{0.5cm}
\caption{(a) Schematic presentation of the spin waves in an elemental ferromagnet.
(b) Schematic presentation of the ground state magnetic structure of a two sublattice
AFM. Sublattices are marked by letters A and B. (c) Schematic presentation of the 
spin waves in a two sublattice AFM. Angles $\theta_A$ and $\theta_B$ 
give deviations of the atomic moments of the sublattices from the magnetization axis. 
Upper part: magnon of type A with $\theta_A>\theta_B$. Lower part:
magnon of type B with $\theta_B>\theta_A$.
}
\label{Fig_magn_structure}
\end{figure}
The structure of a magnon is characterized by the wave vector
and by the angles specifying the deviation of the atomic moments from the magnetization axis.
For an elemental ferromagnet (FM) with one magnetic atom per the crystallographic unit cell the structure of the magnons 
is uniquely determined by the wave vector [Fig.~\ref{Fig_magn_structure}(a)].
The deviation angle $\theta$ is the same for all atoms.
Therefore in this case only the energies of the magnons need to be determined. 
In more complex systems such as antiferromagnets (AFM) [Fig.~\ref{Fig_magn_structure}(b),(c)], 
ferrimagnets, or ferromagnets with several 
inequivalent atoms in the crystallographic unit cell,
different atomic moments deviate differently from the magnetization axis and 
the determination of the structure of the magnons is an important part of the magnon study.  

There are different approaches to the DFT based calculation of magnons.
One of the approaches is the calculation of the dynamic magnetic susceptibility in nonuniform transversal
magnetic field \cite{Savrasov1998,Buczek2011,Sandratskii2012,Odashima2013,Lounis2015,Skovhus2021,Liu2023}. 
A strong feature of this method is that it allows to investigate not only the structure
and the energy of magnons but also their life time resulting from the interaction of magnons with single-electron 
Stoner excitations. In the case of complex magnetic materials the method is very demanding
with respect to both computation techniques and computer resources.
A widely used approach to the theoretical study of the adiabatic magnons is the 
mapping, as an intermediate step, of the electron system on the Heisenberg Hamiltonian of interacting
atomic moments. 
Such a mapping is currently a standard procedure consisting in 
the DFT based evaluation of the Heisenberg exchange 
parameters (see, e.g.,  Refs.~\onlinecite{force_theorem,Mankovsky_parameters2022,
Turek_parameters2006,Lezaic2006,Mu2019,Martinez2023,Sandratskii2007,Sodequist2024}).
An efficient method of the mapping was suggested by Liechtenstein et al. \cite{force_theorem} and is
based on the evaluation of the variation of the band energy as the response to the
infinitesimal deviation of the atomic moments from the magnetization axis. The
possibility to replace the variation of the total energy by the variation of the 
band energy is based on so-called magnetic force theorem \cite{force_theorem}.
A recent review of the development and applications of the Liechtenstein et al. method
is given in Ref.~\onlinecite{Szilva2023}.
In this paper we suggest a direct DFT-based method for the magnon calculation
that does not include mapping of the 
electron system on the Heisenberg Hamiltonian. 
\textcolor{black}{
To our best knowledge this is the first method allowing direct DFT calculations of adiabatic
magnons in complex collinear magnets.
}
The method has important features not provided by the standard mapping approach.
Among them are the following:
First, the fully self-consistent calculation of the magnon structure and energy is performed in contrast
to the mapping approach based on the evaluation of the band energy variation
from non self-consistent calculation.
Second, in the magnon states of complex structures the atoms that are equivalent in the ground state
may become inequivalent. The suggested method takes this into account in a consequent 
self-consistent manner.
Third, in the compounds the contribution of the nominally nonmagnetic atoms
to the magnon states is self-consistently taken into account. 
Forth, 
the method allows to  estimate
the dependence of the magnon energy
on the number of magnons by varying the tilting angles of the atomic moments 
from the magnetization axis.
The approach suggested by Liechtenstein et al. considers an infinitesimal deviation 
of the moments from the magnetization axis.
\textcolor{black}{
Fifth, because of the exact treatment of the generalized periodicity 
of the helical magnetic structures with arbitrary wave vectors our method 
accounts for the exchange interactions between atoms at arbitrary large distances in contrast to 
the mapping calculations that often take into account only a few nearest neighbor
interactions.
}

Since each magnon presents a different state of the system, 
the direct self-consistent calculation of a magnon state
within the DFT framework must include constraints responsible 
for the convergence of the calculation
to the desired magnetic state instead of the 
standard DFT convergence to the ground state (GS).
Two different constraints are simultaneously used in the method. The first governs the change of the 
net magnetization. The method uses the formula 
\begin{equation}
\omega=\frac{\Delta E}{\Delta m_z}  
\label{eq_omega}
\end{equation}
derived in Ref.~\onlinecite{Niu99} for the magnon energy in an arbitrary collinear 
magnet\textcolor{black}{\cite{comment_Niu}.
}
Here  $\Delta E$ is the increase of the energy of the magnon state with respect
to the GS and $\Delta m_z$
is the magnetization change with respect to the GS magnetization. 
Equation~(\ref{eq_omega})  correlates with the property 
that one magnon changes the magnetization of the system by 1~$\mu_B$. 
The first constraint imposes the condition on the magnetization of the system
by means of introduction of an effective magnetic field.

The second constraint specifies the value of the magnon wave vector.
This is a symmetry type of constraint \cite{Sandratskii2001} where the invariance of the initial Hamiltonian with respect 
to the symmetry operation responsible for the desired property reproduces itself during iterations. 
The symmetry constraint does not 
need a constraining field. It is important to emphasize that to constrain the 
magnon wave vector the generalized translational periodicity described by the machinery of 
spin space groups (SSG) must be imposed.
\textcolor{black}{
The combination of two different constraints is the characteristic feature of the method.
}

\begin{figure}[t]
\includegraphics*[width=6cm]{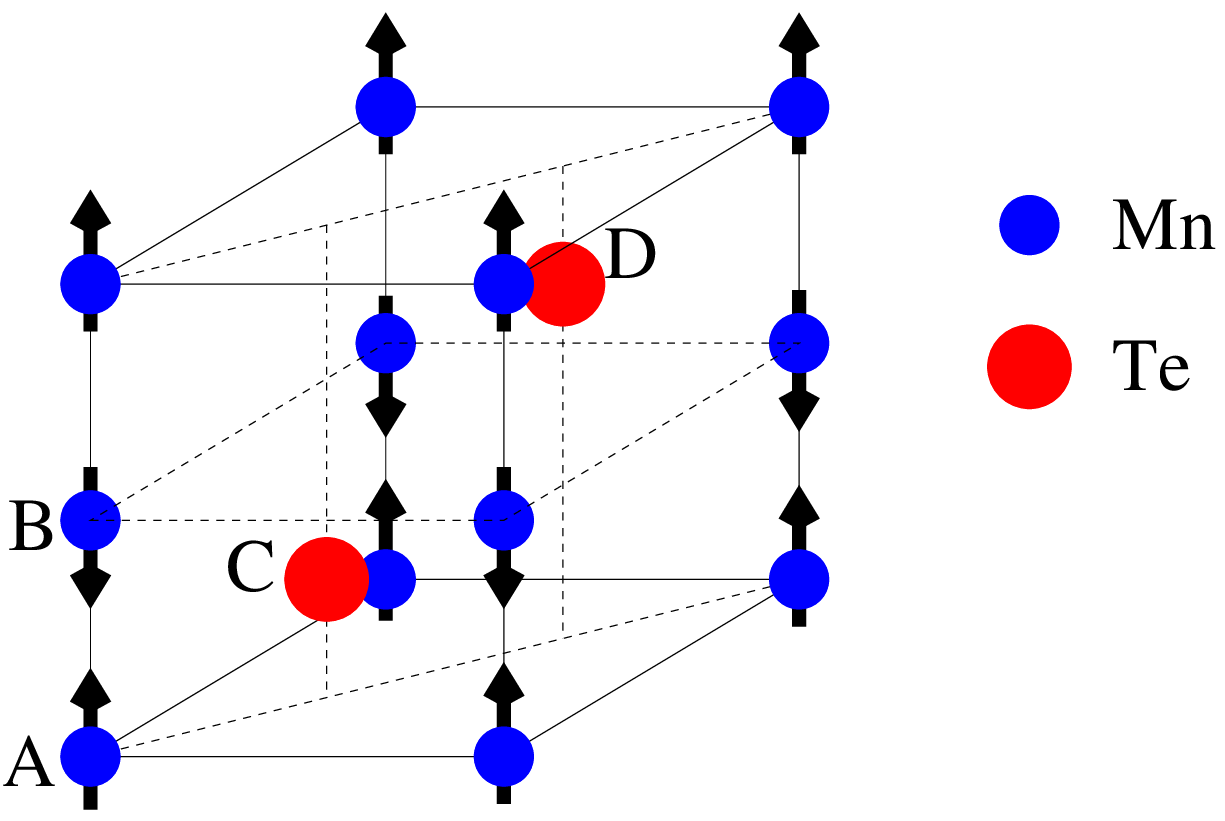}
\caption{Unit cell of AFM MnTe. Labeling A-D of the sublattices is used throughout the paper.}
\label{fig_unit_cell}
\end{figure}
In the paper, we formulate the method in the form  
valid for the study of the magnons in any collinear magnet 
without
spin-orbit coupling (SOC). The performance
of the method is demonstrated by the application to altermagnets focusing on MnTe
as a representative of this class of materials. 
As the notion of an altermagnet is rather new it is worth to introduce it briefly.
The ground state of a two sublattice AFM is characterized by two
mutually compensating ferromagnetic sublattices.  
In terms of the electron structure, the zero net magnetization 
of an AFM is the result of the mutual compensation of the spin-up and spin-down electron states
which is the consequence of the symmetry properties of the material. Importantly, the realization
of this spin compensation can be different for different AFM materials: It
can take place either at each 
wave vector $\bf{k}$ of the reciprocal space
or, alternatively, between the electron states corresponding to different $\bf{k}$ points. 
\v{S}mejkal,  Sinova, and Jungwirth \cite{Smejkal2002,Smejkal2002A} suggested the term 
altermagnet for the materials where the magnetic compensation takes place
between different $\bf{k}$ points. The absence of the spin degeneracy at the same $\bf{k}$ point can be 
treated as the spin splitting
of the electron states at this point. This property of altermagnets  has important physical 
consequences attracting 
\textcolor{black}{
enormous
}
research attention to this class of 
materials\cite{Smejkal2002A,Mazin2002,Bai2024,Reimers2024,Krempasky_MnTe_2024,Mazin2023,fedchenko2024}.

Among the special properties of altermagets is the chirality splitting of the magnon 
states \cite{Smejkal2023,Bai2024,Sodequist2024,Liu2024,Alaei2025}.
\textcolor{black}{
In Ref.~\onlinecite{Smejkal2023} the presence of the magnon chirality splitting is 
connected with the properties of the Heisenberg
exchange parameters. A general approach to the chirality splitting applicable in the DFT-based
direct calculations has not yet been suggested. Also the discussion of the symmetry pattern of the
magnon chirality splitting in the reciprocal wave-vector space and its relation to the symmetry
partern of the exchange splitting in the ground-state band structure is not available.
To give the answers to these questions is one of the aims of this paper.
}

Each AFM magnon brings either positive or negative magnetization to the system. 
If in Fig.~\ref{Fig_magn_structure}(c) $\theta_A$$>$$\theta_B$ the magnon gives negative
contribution to the magnetization whereas for a magnon with $\theta_A$$<$$\theta_B$ the
contribution to the magnetization is positive. These two types of magnons have opposite
chiralities. The two types of magnons are obtained in both Heisenberg model (see, e.g., Ref.~\onlinecite{Rezende2019})
and first-principles calculation of the dynamical spin susceptibility (see, e.g., Ref.~\onlinecite{Sandratskii2012}).
We will refer to these two types of magnons as magnons of type A and B according to the larger of two angles 
$\theta_A$ and $\theta_B$.
The net magnetization of an AFM at nonzero temperatures remains zero.
This property is the consequence of the mutual compensation
of the magnons of opposite chiralities which results from their symmetry-determined energy degeneracy. 
Again there are two possibilities. The chirality compensation can take place either at each magnon wave vector $\bf{q}$
or only between magnons with different wave vectors. The latter situation takes place in altermagnets
and may be referred to as chirality splitting of magnons with a given wave vector $\bf{q}$.
The nature of the chirality degeneracy 
of the magnons is very different compared to the spin degeneracy 
of the electron states discussed above. 
However, in both cases the two symmetry questions to address
are similar: First, which symmetry operations are responsible for the degeneracy 
of the magnons with opposite chiralities, and, second, does the degeneracy 
take place between the magnons with the same wave vector or with different wave vectors? 
As will be demonstrated, the answers to these questions for magnons are closely related to those for 
the electron states. 

Our choice of altermagnet MnTe as the object of the application of the method has following
reasons. First, as a particular case of a two-sublattice AFM containing nonmagnetic atoms 
(Fig.~\ref{fig_unit_cell})
it is complex enough to demonstrate important features of the method.
In the case of MnTe, the
altermagnetic properties are the consequences of the presence of the Te atoms. 
Indeed, in an assumed material with removed Te atoms 
the altermagnetic spin and chirality splittings are absent. 
\textcolor{black}{
Due to the crucial role
of Te atoms in altermagnetism, accounting for their self-
consistent response to changes in the Mn subsystem is
a key aspect. In this context, key questions to address
include the questions of whether the Te atoms remain equivalent in the
magnon states and of what symmetry arguments can reveal
about the induced Te moments. 
}
\textcolor{black}{The magnon-specific symmetry information about nonmagnetic atoms, besides its 
general physical importance,
helps to control and accelerate the convergence of the magnon calculation.
These 'technical' advantages from the symmetry analysis are discussed in Sec.~\ref{sec_sublattices}.}

Next question addressed in the paper is how the chirality properties of the magnon states of the system are related to
the properties of the electron band structures of the magnon states. 
\textcolor{black}{
To our best knowledge this is the first study of this type.
It is based on the possibility to calculate the band structure of the magnon states
opened by our method.
}
The deep connection between two different energy characteristics of altermagnets, 
\textcolor{black}{
magnon energies and magnon electron band structures, 
}
is exposed.

As seen from the above, the symmetry aspects play in the paper an important role.
There are three different parts of the work 
where the symmetry arguments are essential: 
(i) the formulation of the method of the direct magnon calculation,
(ii) the study of the spin-splitting of the electron states in the GS of an altermagnet, and (iii) the study of the 
chirality-splitting of the spin waves in 
an altermagnet. The employment of the SSG allows both solving these tasks and 
the integration of different parts of the study in one coherent physical picture.
\textcolor{black}{
We consider this coherent picture uniting very different sides of the consideration as one of important
results of the paper. As discussed below, in part (ii) the employment of usual space groups can be technically 
sufficient. In other 
parts and in the formation of a general picture the use of the generalized spin-space groups is 
essential.
}

The paper is structured as follows. In Sec.~\ref {sec_method} the method of direct 
DFT based calculation of magnon states is presented. Section~\ref{sec_calc_details}
gives the details of the calculations. In Sec.~\ref {sec_sym_and_results} the symmetry
aspects and the results of the calculations are discussed. This section includes
a brief introduction of the SSGs (Sec.~\ref{sec_SSG}), discussion of the spin degeneracy
and altermagnetic spin-splitting of the electron states in AFM  (Sec.~\ref{sec_degeneracy_electron_states}),
application to the electron states of MnTe (Sec.~\ref{sec_electr_states_MnTe}),
discussion of the symmetry governed properties of the magnetic sublattices in the magnon states
(Sec.~\ref{sec_sublattices}) and of the chirality splitting of magnons in altermagnets  (Sec.~\ref{sec_chirality_symmetry}),
results of the calculation of magnon dispersion (Sec.~\ref{sec_magnon_calculat}),
brief discussion of the noticed instability cases of magnon state energy with respect to the number of magnons
(Sec.~\ref{sec_singular}), and the study of the relation between electronic band structures of the magnon states 
and chirality properties of these states (Sec.~\ref{sec_trans_path}). 
Section~\ref{sec_conclusions} is devoted to the conclusions.

\section{The method of direct DFT-based magnon calculation}
\label{sec_method}
 
As mentioned in the introduction, in Ref.~\onlinecite{Niu99} it was shown that for any 
collinear magnet the magnon energy can be 
\textcolor{black}{
presented
within the framework of the DFT theory
in the form given by Eq.~(\ref{eq_omega}) (see also Ref.~\onlinecite{comment_Niu}).
}
The value of $\Delta m_z$ corresponds to
the number of magnons with given wave vector and $\Delta E$ is their energy.
It is expected that there is an interval of $\Delta m_z$
where $\Delta E$ is proportional to $\Delta m_z$ and 
the calculation for any $\Delta m_z$ from this interval 
gives the value of the spin wave energy. The both quantities can be considered per unit cell.
To obtain the self-consistent magnon state with a given magnetization ${m}_\circ$$-$$\Delta  m_z$ 
the minimization of the density functional must be performed under the constraining 
condition
\begin{equation}
\int d{\bf r}  m_z({\bf r}) =  {m}_\circ - \Delta  m_z.
\label{eq_magn_constr}
\end{equation}
Here ${m}_\circ$ is the ground state magnetization that is zero in the case of an AFM.

\begin{figure}[t]
\includegraphics*[width=6cm]{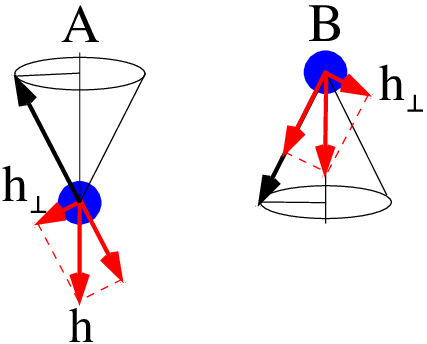}
\caption{Schematic picture of the decomposition of the constraining field {\bf h} into two components: 
one collinear to the moment and the other,  {\bf h}$_\bot$, orthogonal to it. The left (right) part of the figure 
shows the decomposition
for atoms of sublattice A (B). The field is antiparallel to the $z$ axis and leads to the magnon state assigned to
sublattice A. For magnons of type B the direction of field {\bf h} is opposite. 
}
\label{fig_constr_field}
\end{figure}

The constrained energy functional takes the form
\begin{equation}
E_{const}[n,{\bf m}]=E[n,{\bf m}]+
h\left[\int d{\bf r}\:  m_z({\bf r}) -  {m}_\circ + \Delta  m_z\right]
\label{eq_funct_constr}
\end{equation}
where $E[n,{\bf m}]$ is unconstrained functional, Lagrange parameter 
$h$ plays the role of the $z$ component of an effective magnetic field ${\bf h}=(0,0,h)$.
The condition on the magnetization [Eq.~(\ref{eq_magn_constr})] is the same for all magnons
independent of their wave vectors whereas the value of field $h$ corresponding to a given 
$\Delta m_z$ is $\bf{q}$ dependent.
In systems with well defined atomic moments considered in the paper,  field $\bf{h}$ governs the values of the
deviations of the moments from the $z$ axis. 
For each magnetic atom, the vector of the constraining field $\bf{h}$ can be decomposed into two components: 
one collinear to the moment and the other
orthogonal to it (Fig.~\ref{fig_constr_field}). The component collinear to the moment influences the value 
of the moment. This influence is weak for well defined atomic moments since the 
variation of the value of the moments is energetically costly.
In this case the orthogonal component plays the main role governing the deviation 
of the moments from the magnetization axis and leading to the desired magnetization change $\Delta m_z$. 
The orthogonal components form nonuniform magnetic field with wavelength identical to the wavelength
of the magnon. 

The second constraint specifying the wave vector $\bf{q}$ of the magnon
reflects the generalized periodicity \cite{adv} of the helix with given $\bf{q}$  
\begin{equation}
\{\alpha_{{\bf q}n}|E|{\bf R}_n\}{\bf m}({\bf r}) 
\equiv \alpha_{{\bf q}n}{\bf m}({\bf r}-{\bf R}_n)
={\bf m}({\bf r}) 
\label{eq_sym_constr}
\end{equation} 
Here 
$\{\alpha_{{\bf q}n}|E|{\bf R}_n\}$ are the operators of
generalized translations consisting from the lattice translation ${\bf R}_n$
accompanied by the rotation of magnetization by angle ${\bf q}{\bf R}_n$
about the $z$ axis. $E$ in the second position in the symmetry operator
means that the operator does not perform any point transformation
besides the magnetization rotation.
The Kohn-Sham equation of a helical structure in external field {\bf h} takes the form
\begin{equation}
\hat{{\bf H}}\left(\begin{array}{c}
\psi_+\\ \psi_- \end{array}\right) = E \left(\begin{array}{c}
\psi_+\\ \psi_- \end{array}\right)
\label{eq_KS_equation}
\end{equation}
with Hamiltonian
\begin{equation}
\begin{split}
\hat{{\bf H}}=& \hat{{\bf T}}\left(\begin{array}{cc} 1&0\\0&1\end{array}\right)\\
&+ \sum_{n\nu}
    {\bf U}^{+}(\theta_{\nu},\phi_{n\nu})
    \, {\bf V}_\nu({\bf r}_{n\nu})
    \, {\bf U}(\theta_{n\nu},\phi_{n\nu})
    +
    h\sigma_z
\end{split}
\label{eq_hamiltonian}    
\end{equation}
where  $\hat{{\bf T}}$ is the operator of kinetic energy,   $n$ numbers unit cells, $\nu$
numbers atomic sublattices, ${\bf V}_\nu$ is two by two 
potential of the $\nu$th atom in the local atomic spin coordinate system,
${\bf r}_{n\nu}$$=$${\bf r}$$-$${\bf a}_\nu$$-$${\bf R}_n$,
${\bf a}_\nu$ gives the position of the $\nu$th atom in the unit cell,
angles $\theta_{\nu}$ and $\phi_{n\nu}$ specify the direction of the
moment of the $n\nu$th atom,
$ \phi_{n\nu} = \phi_\nu + {\bf q}{\bf R}_n $,
$ {\bf U}(\theta_{n\nu},\phi_{n\nu}) $ is the standard
spin-$ \frac{1}{2} $-rotation matrix, and $\sigma_z$ is the Pauli matrix.
The action of the generalized translation on the spinor wave function 
$\Psi({\bf r})=\left(\begin{array}{c}
\psi_+({\bf r})\\ \psi_- ({\bf r})\end{array}\right) $
has the form
\begin{equation}
\{\alpha_{{\bf q}n}|E|{\bf R}_n\}\Psi({\bf r})=
 {\bf U}(\alpha_{{\bf q}n})\Psi({\bf r}-{\bf R}_n).
\label{eq_genTransl_on_spinor} 
\end{equation}
The Hamiltonian  
of a {\bf q}-magnon [Eq.~(\ref{eq_hamiltonian})] commutes with generalized translations 
corresponding to given {\bf q} \cite{Sandratskii1986,Sandratskii1991}.
The symmetry properties of the Hamiltonian govern the symmetry properties of 
the calculated electron states that
leads to the reproduction of the 
generalized translational symmetry of the Kohn-Sham Hamiltonian during iterations.
Because of this, the fulfillment of the symmetry constraint [Eq.~(\ref{eq_sym_constr})] does not require
application of a constraining field. 

The symmetry with respect to the generalized translations fulfills the conditions of the generalized
Bloch theorem and allows an exact reduction of the calculation for the spiral structure with
arbitrary wave vector to the consideration of the small crystallographic
unit cell of the crystal \cite{Sandratskii1986}.

\section{Calculation details}
\label{sec_calc_details}

The calculations were performed with the augmented spherical waves (ASW) method \cite{Williams1979,Eyert2012}. 
The local density
approximation (LDA) to the exchange-correlation functional
is used\cite{barth72}. 
The account for generalized translation
symmetry and external magnetic field have been implemented earlier \cite{Uhl1992,Miyake2018}. 
Therefore, only limited adaptation of the code were needed.
\textcolor{black}{
}
\begin{figure}[t]
\includegraphics*[width=4cm]{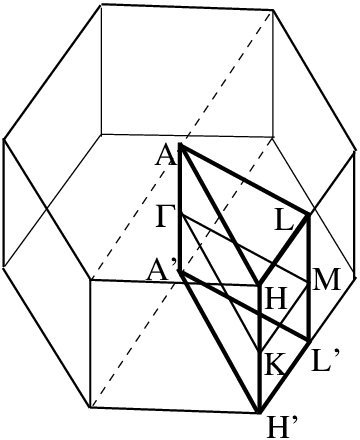}
\caption{Brillouin zone and irreducible domain for calculation of both 
electronic band structure of the AFM GS in the {\bf k} space and spin waves in the {\bf q} space.}
\label{fig_BZ}
\end{figure}
The calculations were performed with three different {\bf k}  meshes in the Brillouin zone (BZ)  
of the material (Fig.~\ref{fig_BZ}): 
$n$$\times$$ n$$ \times n$, $n$=10, 20, and 30.
Here $n$ is the number of the intervals in which the primitive vectors of the reciprocal lattice were divided. The results
of the calculations have shown that the difference between spin wave energies obtained with
$n$=20 and $n$=30 is usually small and most of the calculations reported in the paper were performed 
with $n$=20. 
\textcolor{black}{
In the calculations we used experimental lattice parameters\cite {Szuszkiewicz2006} 
$a=$~4.15~{\AA}, $c=$~6.71~{\AA}. 
}

\begin{figure}[t]
\includegraphics*[width=8cm]{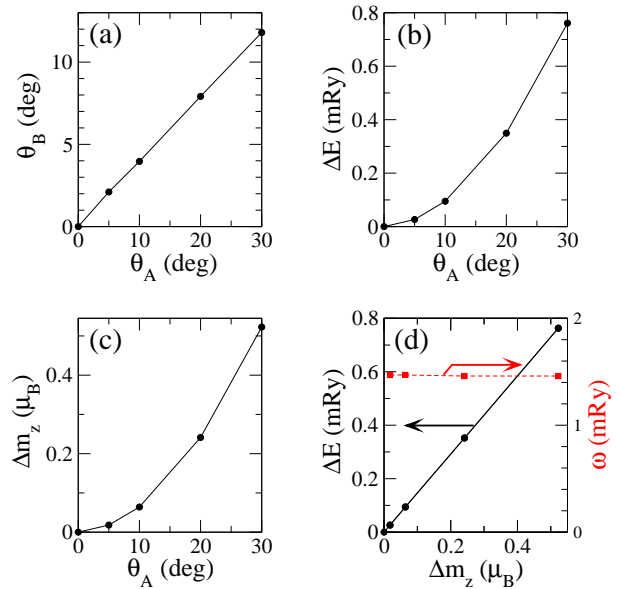}
\caption{The dependence of $\Delta E$ on $\Delta m_z$. Figure shows calculations
performed for magnon with wave vector {\bf q}=(0,0,0.3).
(a)-(c) The dependences of, respectively, $\theta_B$, $\Delta E$, and  $\Delta m_z$ on $\theta_A$. 
(d) The dependence of energy increase  $\Delta E$ (black circles, left energy scala) and magnon energy 
$\omega$  (red squares, right energy scala) on $\Delta m_z$. 
\label{Fig_SW_EofTheta}
}
\end{figure}
As mentioned above, a straightforward application of Eq.~(\ref{eq_omega}) assumes the existence of an interval 
of $\Delta m_z$ where energy increase $\Delta E$
is proportional to magnetization change $\Delta m_z$. Our calculations confirmed that usually 
such a linear dependence exists up to
rather large deviation angles of the Mn atomic moments from the magnetization axis. Figure~\ref{Fig_SW_EofTheta} 
shows typical characters of the dependences between calculated quantities and demonstrates high stability of the 
calculated magnon energy with respect to the
value of the deviation of the Mn moments from the magnetization axis. The results of the calculations
presented in the figure were obtained for wave vector {\bf q}=(0,0,0.3). 
Here and in the rest of the paper we give the $x$ and $y$ components of the reciprocal space vectors 
in units of $\frac{2\pi}{a}$ and the $z$ components in units of $\frac{2\pi}{c}$.
We calculated constraining field $h$ stabilizing the deviations of the Mn moments of sublattice A at 
following values: $\theta_A$=5$^\circ$,10$^\circ$,20$^\circ$,30$^\circ$. 
Figure~\ref{Fig_SW_EofTheta}(a) gives calculated deviations of the moments of sublattice B, $\theta_B$. 
Figure (b) shows corresponding 
increase in energy $\Delta E$.
The difference between
$\theta_A$ and $\theta_B$ is the source of the magnetization change $\Delta m_z$ shown in figure (c).
In figure (d) we present the dependence of $\Delta E$ on $\Delta m_z$ which is very close to a linear one.
The ratio of these quantities giving the magnon energy is with a good accuracy independent of  $\Delta m_z$.
In most of the calculations presented in the paper we used  $\theta_A$=20$^\circ$. 
On the other hand, 
\textcolor{black}{ performing model calculations aimed at testing our method  
we obtained also
the cases where strong deviation from the simple dependence of $\Delta E$ on the 
deviation angle of the Mn atomic moments was registered showing the capacity of the method 
to reveal an instability 
connected with the deviation of the atomic moments from the
magnetization axis.
These results are briefly presented in Sec.~\ref{sec_singular}.
}

\section{Symmetry aspects and results of calculations}
\label{sec_sym_and_results}
\subsection{SSG groups}
\label{sec_SSG}

The standard tool for the analysis of the symmetry properties of the crystalline materials is the
apparatus of space groups. For magnetic systems an antisymmetry operation is included
into consideration making extension from space groups to magnetic space groups\cite{Shubnikov1964,Padmanabhan2020}. 
The antisymmetry operation that in an abstract treatment can be considered as changing the color between black and white 
in the magnetic case is responsible for the reversal of the magnetization 
\textcolor{black}{
direction by means of time reversal.
}
However,  in the problems where the influence of the SOC can be neglected this tool is 
not sufficient for the description of the properties of magnetic systems.
In particular, these limitations have been revealed for both
electron band structure calculations and the studies based on the Heisenberg Hamiltonian of interacting 
atomic moments.

The solution of the problem has been found in the concept of SSG whose 
elements allow different transformation of the
spin and space variables (see, e.g., Refs.~\onlinecite{Brinkman1966A,Brinkman1966B,Cracknell1974}). 
In recent years the interest to SSG has been revived.
Reference~\onlinecite{Corticelli_SSG2022} reports a detailed analysis of the magnon band topology 
within the framework of the Heisenberg and Heisenberg-Kitaev models of interacting atomic moments.
Very recently several systematic works\cite{Xiao2024,Cheng2024,Jiang2024} were published devoted 
to the classification, properties, and applications of the SSGs.

The action of the SSG operator $\{\alpha_S|\alpha_R|\boldsymbol{\tau}\}$ 
on the magnetization ${\bf m}({\bf r})$ is defined as 
\begin{equation}
\{\alpha_S|\alpha_R|\boldsymbol{\tau}\}{\bf m}({\bf r}) = \alpha_S {\bf m}([\alpha_R|\boldsymbol{\tau}]^{-1}{\bf r})
\label{eq_SSG_operator}
\end{equation}
where $\alpha_S$ and $\alpha_R$ are spin and space rotations respectively, $\boldsymbol{\tau}$ is 
space translation,
and we introduced notation $[\alpha_R|\boldsymbol{\tau}]$$\equiv$$\{E|\alpha_R|\boldsymbol{\tau}\}$. 
The action of the SSG operator on the two-component spinor has the form
\begin{equation}
\{\alpha_S|\alpha_R|\boldsymbol{\tau}\}\Psi({\bf r})=
{\bf U}(\alpha_S)\Psi([\alpha_R|\boldsymbol{\tau}]^{-1}{\bf r})
\label{eq_SSG_on_spinor}
 \end{equation}
The generalized translations [Eq.~(\ref{eq_genTransl_on_spinor})], crucial for the study of spiral magnetic configurations,
are the operations of the SSG.
The operation of time reversal 
acting on two-component spinor takes the form \cite{} 
\begin{equation}
\Theta=-i\sigma_yK=\left(\begin{array}{cc} 0&-1\\1 &0 \end{array}\right) K 
\end{equation}
where $K$ is the operator of complex conjugation.
Since $\left(\begin{array}{cc} 0&-1\\1 &0 \end{array}\right)$ 
belongs to the set of unitary matrices {\bf U} entering Eq.~(\ref{eq_SSG_on_spinor}), it already belongs to the SSG as an 
operation providing a spin rotation.
Therefore, the complex conjugation $K$ is also an allowed SSG operation. 
This means that the real form of the Kohn-Sham equations of collinear magnets [Eq.~(\ref{eq_hamiltonian})] and its 
consequences become within SSGs a part of a straightforward symmetry treatment.

The effectiveness of applying SSG, compared to space
groups, can be characterized as follows.
The neglect of the SOC leads effectively to the 
replacement of the actual physical 3D space by the  6D space where spin and orbital 
variables are independent and can be transformed
separately. The account for this freedom gives important new information about the properties of the 
system.
In addition, the SSGs allow a straightforward establishment of  continuity relations between theoretical results obtained 
within different approximations,
such as 
with and without SOC, or between the results obtained for collinear and 
helical magnetic configurations. The reason for this is the property that the SSG of the less symmetric case is the 
subgroup of the SSG of the more symmetric case.
The latter feature plays important role in Sec.~\ref{sec_trans_path} 
where we discuss the transformation path of the electron band structure
of a collinear magnet to the electron band structure of a magnon with a given wave vector {\bf q}.

Summarizing the applications in the paper of the symmetry concepts, we distinguish three different problems.
The first 
was considered in Sec.~\ref{sec_method} and uses the generalized translations as a symmetry constraint 
in the calculation of the magnon states. The second and third are, respectively, the 
altermagnetic spin-splitting of the electron states and chirality splitting of the magnon states. 

\subsection{Degeneracy of the electron states in collinear magnets}
 \label{sec_degeneracy_electron_states}
 
\textcolor{black}{
The material of this section contributes to making the paper reasonably self-contained and 
provides the basis for the comparison with the results on the magnon chirality
splitting discussed in Sec.~\ref{sec_chirality_symmetry}.
}

The electronic band structure of collinear magnets has been calculated already  
for distinctly more than 50 years (see, e.g., Refs.~\onlinecite{Wakoh1966,Asano1967}).
The electron wave functions were 
treated as scalar functions labeled with an index specifying the sign of the spin projection
on the selected quantization axis.
Respectively, two scalar Schr{\"o}dinger equations were considered,
one for each spin projection. 
The symmetry-caused degeneracy of the electron states arises in the following way. 
If $g=[\alpha_R|\boldsymbol{\tau}]$  is a symmetry operation commuting with a scalar Schr{\"o}dinger equation, 
the action of this operation on 
an eigenstate $\psi_{{\bf k}\sigma}$ gives the eigenstate with the same energy, the same spin projection $\sigma$, 
and wave vector $\alpha_R\bf{k}$. 
The real form of the equation additionally gives the degeneracy of the electron states at points $\bf{k}$ and $-\bf{k}$.
The different vectors from the list of all $\alpha_R \bf{k}$ and $-\alpha_R \bf{k}$ vectors form the star $\{\bf{k}\}$ 
of vector $\bf{k}$ \cite{comment_star_k}.

In a FM, the Schr\"{o}dinger equations for opposite spin projections are essentially different and, therefore,
there is no symmetry-caused degeneracy of the spin-up and spin-down states. The spin splitting
is the term characterizing this property of the electron structure which is valid for
each point of the {\bf k} space.

On the other hand, in an AFM there must be degeneracy between the states with opposite spin projections.
This degeneracy is the reason for the zero net magnetization of the AFM.
To expose the origin of the spin degeneracy we first notice that in the Schr\"{o}dinger equation for electrons with a
given spin projection $\sigma$ the electrons see different potentials at the atoms of different magnetic sublattices. 
Therefore at the first step only the operations leaving the sublattices invariant are considered. For each vector
$\bf{k}$ they give star $\{{\bf k}\}_{subl}$ corresponding to the symmetry group of the sublattices.  
Next it is necessary to specify 
the symmetry operation that transforms the equations corresponding
to different spin projections into each other. The real-space part of this operation $[\alpha^{tr}_R|\boldsymbol{\tau}]$ 
must transform the
sublattices into each other and be accompanied by an 'antisymmetry' operation $E'$ reversing the 
signs of the spin indices. 
\textcolor{black}{
The physical nature of this antisymmetry operation is time reversal.
}
As a consequence of such symmetry operation any spin-up state at point $\bf{k}$
is degenerate with a spin-down state at $\alpha^{tr}_R\bf{k}$. This makes the system as a whole spin-compensated.
The answer to the question whether the spin compensation takes place at each $\bf{k}$ point of the BZ or only between 
different $\bf{k}$ points depends on the properties of  
operation $[\alpha^{tr}_R|\boldsymbol{\tau}]$ transforming the sublattices into each other. 
If vector $\alpha^{tr}_R\bf{k}$ belongs to star $\{{\bf k}\}_{subl}$ the degeneracy takes place 
at each $\bf{k}$$\in$$\{{\bf k}\}_{subl}$. In the opposite
case we deal with an altermagnet with spin splitting at all points of the star $\{{\bf k}\}_{subl}$.

This approach to the symmetry properties of AFMs allows to reach the description of the
electron structure without reference to the SSG. Its application to MnTe was reported in old
publication Ref.~\onlinecite{Sandratskii1981}. A very detailed and complete discussion of the application of this type of approach
to the altermagnets was recently published by Turek \cite{Turek2022}.
This approach has shortcomings:  Instead of treating the electron wave functions as spinors
it considers them as scalars labeled with a spin index. This complicates the consideration of the influence 
on the electron band structure of the 
SOC or of the noncollinearity of the magnetic configuration where the account for spinor form of the 
electron wave functions is essential. 
Therefore, this approach is not sufficient
for the purposes of this paper where noncollinear incommensurate spiral structures
are in the focus of the consideration.

The application of the SSG to the symmetry analysis of the collinear magnetic structures  
gives additional useful features. First, any spin rotation $\{C_{z\phi}|E|0\}$ about the $z$ axis 
is a symmetry operation of the spinor Kohn-Sham equation [Eqs.~(\ref{eq_KS_equation}),(\ref{eq_hamiltonian})]. 
As the consequence of this symmetry 
the electron eigenfunctions assume one of the two spinor forms 
$\psi({\bf r})\left(\begin{array}{c}1\\0 \end{array}\right)$
or $\psi({\bf r})\left(\begin{array}{c}0\\1 \end{array}\right)$
corresponding to different irreducible representations (IR) of the SSG group \cite{Sandratskii1979}. Therefore,
the spin-indexing of the electron functions is now a straightforward consequence of the symmetry
of the problem and not the property imposed on the basis of additional arguments.
Another consequence of the description of the collinear
magnetic states in terms of SSG is that these states can be treated as spiral structures with arbitrary wave vectors {\bf q}
and deviation angles $\theta$=$0$. Indeed, the group of generalized translations ${\bf T}_{\bf q}$ with any {\bf q} is a subgroup of 
the group ${\bf C}_z$$\times${\bf T} where ${\bf T}$ is the group of space translations $[E|{\bf R}_n]$ 
and ${\bf C}_z$ is the group
of all spin rotations $C_{z\phi}$
about the $z$ axis.
This property will be used in 
Sec.~\ref{sec_trans_path}
to study the transformation path of the electron structure of the collinear ground state into the electron
structure of the magnon state with a given wave vector.

\subsection{Collinear AFM ground state of MnTe}
\label{sec_electr_states_MnTe}
\begin{widetext}
\begin{onecolumngrid}
\begin{table}[h]
\caption{Point symmetry elements.The following notations are used: $E$ is the unity transformation,
$I$ is the space inversion, $C_n^m$ is the rotation by angle $2\pi\frac{m}{n}$. In the column 'axis' 
the unit vectors parallel to the rotation axes are given. The two operations  presented in each row
have the same rotation axes. Columns headed $\alpha_R\bf{r}$ and $I\alpha_R\bf{r}$ give 
the coordinates of the vectors obtained after rotation of vector ${\bf r}=(x,y,z)$.
In columns headed $\boldsymbol{\tau}$,  symbol \checkmark marks the symmetry operations containing non-primitive translation 
$\boldsymbol{\tau}$=(0,0,0.5). 
In the columns headed 'subl', the operations transforming 
the Mn sublattices into themselves are marked with \checkmark. 
For all direct-space vectors, the $x$ and $y$ coordinates are given in units of lattice parameter $a$ and
the $z$ coordinate in units of lattice parameter $c$.
} 
\begin{center}
$\begin{array}{|c|c|c|c|c|c|c|c|c|c|c|c|c|c|c|c|c}
\hline
\hline
N&\alpha_R &axis&\multicolumn{3}{c|}{\alpha_R\bf{r}}&\boldsymbol{\tau}&subl&\:\:\:\:&N&I\alpha_R&\multicolumn{3}{c|}{I\alpha_R\bf{r}}&\boldsymbol{\tau}&subl\\
\hline
1&E        &&x&y&z&-&\checkmark&\:\:&13&I&-x&-y&-z&-&\checkmark\vspace{1mm}\\
2&C_6^1&(0,0,1)&\frac{1}{2}x-\frac{\sqrt{3}}{2}y &\frac{\sqrt{3}}{2}x+\frac{1}{2}y&z&\checkmark&-&\:\:
&14&IC_6^1&-\frac{1}{2}x+\frac{\sqrt{3}}{2}y &-\frac{\sqrt{3}}{2}x-\frac{1}{2}y&-z&\checkmark&-\vspace{1mm}\\
3&C_6^5&(0,0,1)& \frac{1}{2}x+\frac{\sqrt{3}}{2}y &-\frac{\sqrt{3}}{2}x+\frac{1}{2}y&z&\checkmark&-&\:\:
&15&IC_6^5& -\frac{1}{2}x-\frac{\sqrt{3}}{2}y &\frac{\sqrt{3}}{2}x-\frac{1}{2}y&-z&\checkmark&-\vspace{1mm}\\
4&C_6^2&(0,0,1)& -\frac{1}{2}x-\frac{\sqrt{3}}{2}y &\frac{\sqrt{3}}{2}x-\frac{1}{2}y&z&-&\checkmark&\:\:
&16&IC_6^2& \frac{1}{2}x+\frac{\sqrt{3}}{2}y &-\frac{\sqrt{3}}{2}x+\frac{1}{2}y&-z&-&\checkmark\vspace{1mm}\\
5&C_6^4&(0,0,1)& -\frac{1}{2}x+\frac{\sqrt{3}}{2}y &-\frac{\sqrt{3}}{2}x-\frac{1}{2}y&z&-&\checkmark&\:\:
&17&IC_6^4& \frac{1}{2}x-\frac{\sqrt{3}}{2}y &\frac{\sqrt{3}}{2}x+\frac{1}{2}y&z&-&\checkmark\vspace{1mm}\\
6&C_6^3&(0,0,1)& -x&-y&z&\checkmark&-&\:\:
&18&IC_6^3& x&y&-z&\checkmark&-\vspace{1mm}\\
7&C_2    &(1,0,0)& x&-y&-z&-&\checkmark&\:\:
&19&IC_2    & -x&y&z&-&\checkmark\vspace{1mm}\\
8&C_2    &(\frac{1}{2},\frac{\sqrt{3}}{2},0)& -\frac{1}{2}x+\frac{\sqrt{3}}{2}y &\frac{\sqrt{3}}{2}x+\frac{1}{2}y&-z&-&\checkmark&\:\:
&20&IC_2    & \frac{1}{2}x-\frac{\sqrt{3}}{2}y &-\frac{\sqrt{3}}{2}x-\frac{1}{2}y&z&-&\checkmark\vspace{1mm}\\
9&C_2    &(-\frac{1}{2},\frac{\sqrt{3}}{2},0)& -\frac{1}{2}x-\frac{\sqrt{3}}{2}y &-\frac{\sqrt{3}}{2}x+\frac{1}{2}y&-z&-&\checkmark&\:\:
&21&IC_2   & \frac{1}{2}x+\frac{\sqrt{3}}{2}y &\frac{\sqrt{3}}{2}x-\frac{1}{2}y&z&-&\checkmark\vspace{1mm}\\
10&C_2  &(\frac{\sqrt{3}}{2},\frac{1}{2},0)& \frac{1}{2}x+\frac{\sqrt{3}}{2}y &\frac{\sqrt{3}}{2}x-\frac{1}{2}y&-z&\checkmark&-&\:\:
&22&IC_2  &  -\frac{1}{2}x-\frac{\sqrt{3}}{2}y &-\frac{\sqrt{3}}{2}x+\frac{1}{2}y&z&\checkmark&-\vspace{1mm}\\
11&C_2  &(0,1,0)&-x&y&-z&\checkmark&-&\:\:
&23&IC_2 &x&-y&z&\checkmark&-\vspace{1mm}\\
12&C_2  &(-\frac{\sqrt{3}}{2},\frac{1}{2},0)&\frac{1}{2}x-\frac{\sqrt{3}}{2}y &-\frac{\sqrt{3}}{2}x-\frac{1}{2}y&-z&\checkmark&-&\:\:
&24&IC_2  &-\frac{1}{2}x+\frac{\sqrt{3}}{2}y &\checkmark\frac{\sqrt{3}}{2}x+\frac{1}{2}y&z&\checkmark&-&\vspace{1mm}
\label{table1}
\end{array}$
\end{center}
\end{table}
\end{onecolumngrid}
\end{widetext}

In MnTe, the space group of the atomic lattice contains 24 point operations listed in Table~\ref{table1}: 
12 operations of type I transform the Mn sublattices into themselves whereas 
other 12 operations of type II transform them into each other.
In the AFM state, the point operations of type II
remain the part of the symmetry operations in combination with spin-index reversal operation
\textcolor{black}{
(time reversal operation).
}

\begin{figure}[t]
\includegraphics*[width=6cm]{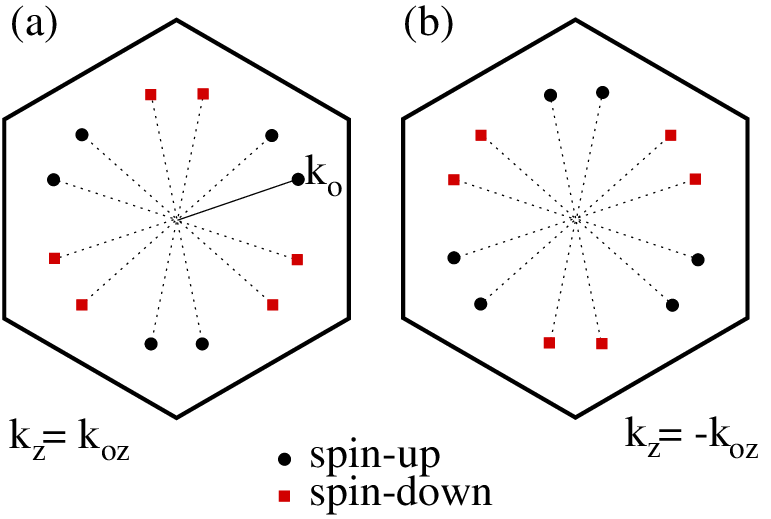}
\caption{Star $\{{\bf k}_\circ\}$ of a general wave vector ${\bf k}_\circ$. The star contains 24 points
lying in two planes $k_z=k_{\circ z}$ [panel (a)] and $k_z=-k_{\circ z}$ [panel (b)]. For any electron state 
at  ${\bf k}_\circ$, the points shown by black circles contain equivalent states with the same spin 
projection. These 12 points form the star $\{{\bf k}_\circ\}_{subl}$ corresponding to the symmetry group 
of the Mn sublattices. The points shown by red squares contain equivalent electron states with opposite spin
projection. At all 24 points there is the altermagnetic spin splitting of the electron states.
As discussed in Sec.~\ref{sec_chirality_symmetry}, the same figures reflect the properties of the 
chirality splitting of the magnons.
In this case the wave vectors {\bf k} of the electron states must be replaced by corresponding magnon 
wave vectors {\bf q} and references to spin projections by references to chiralities.  
\label{Fig_stars}
}
\end{figure}

The application of the 12 operations of type I to a reciprocal-space vector {\bf k}=$(k_x,k_y,k_z)$ gives the
set of 12 vectors with coordinates transformed according to the three columns of Table~\ref{table1} with common 
heading $\alpha_R\bf{r}$ or $I\alpha_R\bf{r}$. For instance, for the operation number 4 the transformed vector is
($-$$\frac{1}{2}k_x $$-$$\frac{\sqrt {3}}{2}k_y$, $ \frac{\sqrt {3}}{2}k_x$$-$$\frac{1}{2}k_y,k_z)$.
These 12 vectors are not necessarily different. 
For the most symmetric 
$\Gamma$ point {\bf k}=$(0,0,0)$ all vectors are equal and the corresponding star 
$\{{\bf k}\}_{subl}$ consists of one vector. 
On the other hand, for a general point   $\bf{k}$ 
all 12 vectors are different.
In Fig.~\ref{Fig_stars} the points of the star $\{{\bf k}\}_{subl}$ are shown for a general vector  
${\bf k}_\circ$=$(k_{\circ x},k_{\circ y},k_{\circ z})$   
and marked as black circles: 6 points lie in the $k_z$=$k_{\circ z}$ plane and
other 6 points in the $k_z$=-$k_{\circ z}$ plane. 
At 12 points of the star $\{{\bf k_\circ}\}_{subl}$ thus obtained,
for each electron state at point ${\bf k_\circ}$ there are equivalent 
states with the same energy and the same spin projection. 

As a representative of the operations of type II we will use the reflection in the $xy$ plane 
(operation number 18 in Table~\ref{table1}).
Action with this operation on the vectors of the star $\{{\bf k_\circ}\}_{subl}$ gives another 12 
vectors shown in Fig.~\ref{Fig_stars} as red squares. 
They differ from the vectors of the star $\{{\bf k_\circ}\}_{subl}$ by the sign of 
the $z$ component.
The states at these 12 vectors are degenerate with the states at
${\bf k_\circ}$ but have opposite spin projection. 
All 24 vectors are different and form full star $\{{\bf k_\circ}\}$ of vector ${\bf k_\circ}$.
 
In general, for any  {\bf k} we have the star  $\{{\bf k}\}_{subl}$ obtained with operation leaving magnetic
sublattices invariant and the full star  $\{{\bf k}\}$ obtained with account for all symmetry operations.
These two sets of vectors are either identical or the number of vectors in $\{{\bf k}\}$ is double.
In the former case, the change of the
sign of the $z$ component of the vectors from $\{{\bf k}\}_{subl}$ does not change the set of vectors and
at all points of the set there is spin degeneracy of the electron states whereas in the latter case 
at all points in $\{{\bf k}\}$ there is the altermagnetic spin splitting. 
 
The irreducible domain of the AFM MnTe is the triangular prism $\Gamma$MKALH that is  
$\frac{1}{24}$th of the BZ (Fig.~\ref{fig_BZ}). 
The analysis of all points of the  irreducible domain shows that the altermagnetic spin splitting takes 
place at all inner points of the prism and, additionally, at inner points of the face  $\Gamma$MLA. 
At these points the calculations 
should be performed separately
for the spin-up and spin-down states. 
Alternatively, the calculations can be performed in the $\frac{1}{12}$th of the BZ, prism ALHA'L'H'  (Fig.~\ref{fig_BZ}),  
but for one spin projection only. 
 
\begin{figure}[t]
\includegraphics*[width=6cm]{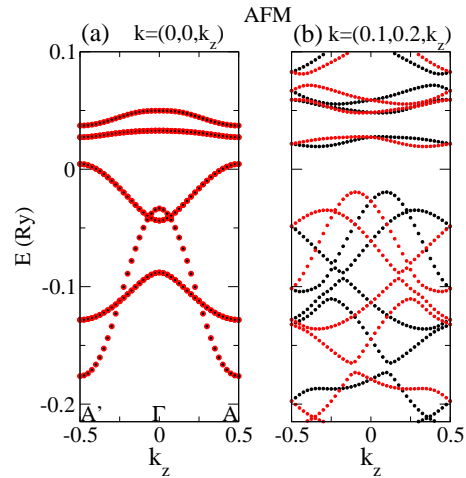}
\caption{Two fragments of the band structure of AFM MnTe. (a) High symmetry line (0,0,$k_z$) in the reciprocal space.
(b) Low symmetry line (0.1,0.2,$k_z$) in the reciprocal space. Black (red) circles present spin-up (spin-down) states. 
At the points of the high symmetry line all states are spin degenerate. In the case of the low symmetry line 
there is altermagnetic spin splitting at all points with exception of the center and end points of the interval. 
\label{Fig_bands_AFM}
}
\end{figure}

In Fig.~\ref{Fig_bands_AFM}  we show one fragment of the calculated band structure with spin-degenerate bands and 
one fragment with altermagnetic spin-splitting.
 
\subsection{Magnetic structure of sublattices in magnon states}
\label{sec_sublattices}
Let us discuss the properties of the structure of the AFM magnons
that follow from the symmetry arguments.
\textcolor{black}{
These results provide important general knowledge about properties of magnons. 
Simultaneously, they can be used in setting up and control of the DFT-based calculations.
For example, the correlation in the directions of the atomic moments 
of different atoms predicted by symmetry must be present in the self-consistent 
solution for the magnon state that provides a tool for controlling the calculations. 
On the other hand, such a correlation can be used from the beginning of calculations
to decrease the number of degrees of freedom and simplify the process of
convergence. In addition, the symmetry-based knowledge that the magnetic atoms become 
inequivalent in the magnon excitations helps to avoid an unphysical assumption
of the preserved equivalence of atoms and an erroneous constraint of the equivalence 
of magnetic atoms in the calculations.
}

We consider the magnon with an arbitrary wave vector {\bf q}. The magnetic structure of the magnon
is invariant with respect to the generalized translations corresponding to given {\bf q}. 
As stated above, the direct magnon calculation includes an effective magnetic field acting on both 
magnetic sublattices. This field is collinear to the $z$ axis and therefore
parallel to the magnetization of one sublattice and antiparallel to the
magnetization of the other sublattice\cite{comment_sign_h}. Hence the atoms of the two sublattices
in the magnon state become inequivalent. 
On the basis of these conclusions, the 
magnetic configurations of the two sublattices in a magnon state can be written in the following form
\begin{widetext}
\begin{align}
&{\bf m}_{nA}=m_A\{\sin(\theta_A)\cos[{\bf q}({\bf a}_A+{\bf R}_n)],\sin(\theta_A)\sin[{\bf q}({\bf a}_A+{\bf R}_n)],
\cos(\theta_A)\} \label{eq_magnon_A} \\
\label{eq_magnon_B} 
&{\bf m}_{nB}=m_B\{\sin(\pi-\theta_B)\cos[{\bf q}({\bf a}_B+{\bf R}_n)+\pi+\phi],\sin(\pi-\theta_B )
\sin[{\bf q}({\bf a}_B+{\bf R}_n)+\pi+\phi],-\cos(\theta_B)\}.
\end{align}
\end{widetext}
Because of the inequivalence of the sublattices
it is expected that $m_A$$\neq$$ m_B$, $\theta_A$$\neq$$\theta_B$, and there is an unknown
not symmetry-governed angle $\phi$ 
specifying the phase shift between $xy$-projections
of the moments of the two sublattices.
The calculations with the suggested method confirmed
that $m_A$$\neq$$ m_B$ and $\theta_A$$\neq$$\theta_B$.
More details on these quantities will be given below. 

However, concerning angle $\phi$ the results of the calculations were unexpected:
iterations started with an arbitrary selected $\phi$$\neq$0
resulted in the self-consistent magnetic configurations with $\phi$=0. 
The question arises why the orientations of the moments of two inequivalent
sublattices not connected by any symmetry operation are in such a
strict coordination with each other.
This type of the coordination is expected to be a symmetry-caused property. The explanation for this 
property is the following. The cone spiral structure
of sublattice A  [Eq.~(\ref{eq_magnon_A})] for arbitrary {\bf q} and $\theta_A$ is invariant with respect to the SSG operation 
$\{\Theta C_{2y}|I|0\}$ that combines space inversion $I$ and spin transformation $\Theta C_{2y}$
performing spin reflection in the $xz$ plane. This operation leaves invariant also
the cone structure of the sublattice B [Eq.~(\ref{eq_magnon_B})] but only in the case of $\phi$=0 or $\phi$=$\pi$. 
This means that if the calculation is started with $\phi$=0
the value of $\phi$ remains zero during iterations since the symmetry with respect 
to the operation $\{\Theta C_{2y}|I|0\}$ must be preserved. 
Thus, 
although there is no symmetry operation transforming the sublattices into each other there is an operation that is
responsible for preserving $\phi$=0 and makes the two-sublattice magnetic configuration distinguished by 
an additional symmetry compared 
with the configurations obtained by a nonzero relative phase shift $\phi$ between the sublattices.
Hence, the choice of $\phi$=0 in the starting magnetic configuration imposes another symmetry 
constraint \cite{Sandratskii2001}  in the magnon calculation that is additional to the constraint of generalized periodicity.
 
The presence of the
symmetry operation 
effectively connecting inequivalent Mn sublattices 
influences also the properties of the Te atoms in the magnon states. 
In contrast to the Mn sublattices, the atoms of the two Te sublattices remain equivalent  
because the space inversion
transforms the Te sublattices into each other. This operation influences also the directions
of the induced Te moments since to keep this symmetry operation intact
the induced atomic moments of the two Te sublattices must transform into each other.
The straightforward symmetry analysis gives for the Te atoms in the unit cell 
\begin{align}
\label{eq_Te_theta}
&\theta_C=\theta_D\equiv\theta_{Te}\neq0\\ 
\label{eq_Te_phi}
&\phi_C+\phi_D={\bf qR}_{IC}
\end{align}
where lattice vector
${\bf R}_{IC}$=${\bf a}_D$$-$$I{\bf a}_C$.
Although the values of angles $\theta_C,\theta_D$ and $\phi_C,\phi_D$ cannot be determined by symmetry arguments 
the knowledge of relations~(\ref {eq_Te_theta}),(\ref {eq_Te_phi}) allows to
accelerate the convergence process for the magnon configurations by adequate preparation of the
initial magnetic configurations.
\textcolor{black}{
The calculation process includes the self-consistent determination of the 
directions of all atomic moments. This process can be rather time consuming. 
The knowledge of the symmetry relations between the moments of different atoms
allows to decrease the number of the degrees of freedom and leads to a considerable reduction
of the convergence time.
}

Here we give some further information concerning the calculated values of sublattice quantities 
entering Eqs.~(\ref{eq_magnon_A}),(\ref{eq_magnon_B}).
The values of the Mn moments appeared to be very robust and 
the difference between $m_A$ and $m_B$ was always small: for calculations with $\theta_A$=20$^\circ$ it
never exceeded a few thousandth of Bohr magniton for estimated Mn atomic moments of $\sim$4.2$\mu_B$. 
On the other hand, the difference between $\theta_A$ and $\theta_B$ is an essential feature of the 
magnon states. It is strongly {\bf q}-dependent.

For field {\bf h} antiparallel to the $z$ axis $\theta_A$$>$$\theta_B$. Such magnon states have negative magnetization
with respect to the z axis and is associated with sublattice A.
For field {\bf h} parallel to the $z$ axis 
$\theta_A$$<$$\theta_B$, the magnetization is positive, and magnon is associated with sublattice B.
The magnons associated with different sublattices have opposite chiralities. 

 %PUT SOMEWHERE
\begin{figure}[t]
\includegraphics*[width=2.5cm]{ 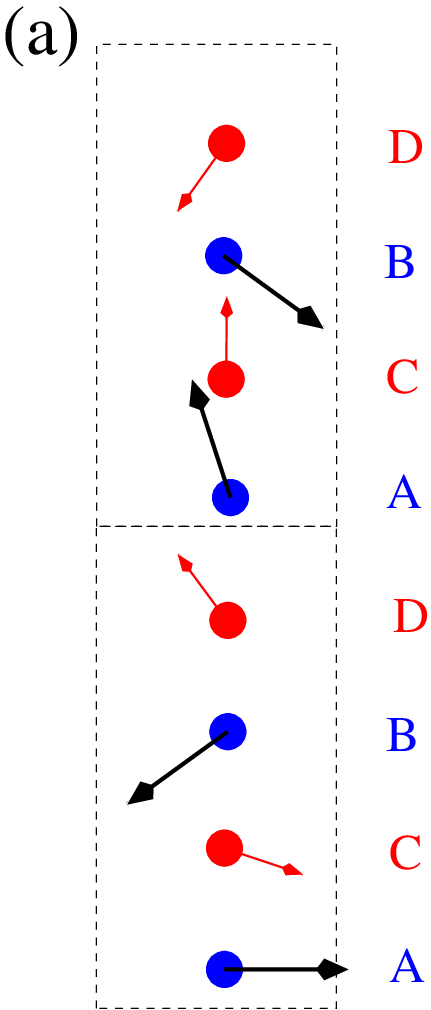}\hspace{0.5cm}
\includegraphics*[width=3cm]{ FIG10_Delta_phi.eps}
\caption{(a) Directions of the projections of the atomic moments on the $xy$ plane
calculated for {\bf q}=(0,0,0.3) and $\theta_A$=$20^\circ$. The atomic labeling is according
to Fig.~\ref{fig_unit_cell}. Two unit cells adjacent along the $z$ axis are presented. 
(b) For each atom we show the value of the angle between the $xy$ projection of its
moment and of the moment of the atom lying next below. 
}
\label{fig_phi}
\end{figure}

 For completeness we give some results of the calculation of the values and directions of the 
induced Te moments. The both characteristics are {\bf q} dependent. The induced moments deviate 
rather weakly from the $z$=0 plane. For 
magnons of type A, $\theta_{Te}$$>$90$^\circ$ and the $z$ component of the Te moments is
negative. Respectively,  for 
magnons of type B, $\theta_{Te}$$<$90$^\circ$ and the $z$ component of the Te moments is
positive. If the constrained deviation of the Mn moment is 20$^\circ$, the maximal deviation of the Te moments 
from the $z$=0 plane is about 20$^\circ$. The values of the induced moments in this case do not
exceed a few thousandth of Bohr magniton. Therefore the contribution of the Te moments to the 
magnetization of the magnon states that is collinear to the $z$ axis is weak. On the other hand, 
the contribution of the induced moments into magnon energy can be noticeable.  For chirality 
degenerate magnons of A and B type, the values of the Te moments and of their deviations from
the  $z$=0 plane are exactly equal. For the chirality split magnons at a given {\bf q} the values are 
numerically different though in MnTe this difference is not large.

In Fig.~\ref{fig_phi}(a) we show the directions of the projections of the moments on the $xy$ plane
calculated for {\bf q}=(0,0,0.3) and $\theta_A$=$20^\circ$. Two unit cells adjacent along the $z$ axis are presented. 
The directions of the atomic moments in the second unit cell rotate by angle 
{\bf qR} with respect to the corresponding moments in the first unit cell.
Here {\bf R} is the lattice vector connecting the cells.
The relative directions 
of the Mn moments of sublattices A and B in the same unit cell are also according 
to wave vector {\bf q}: $\phi_B$-$\phi_A$= ${\bf q}({\bf a}_B$-${\bf a}_A)$+$\pi$
that is the consequence of $\phi$=0 in Eq.~(\ref{eq_magnon_B}).
The directions of the Te moments (atoms C and D) within one unit cell cannot 
be uniquely determined by symmetry. There is, however, symmetry property of 
equal angles of the Te moments with respect to the moment of the Mn atom lying 
between them. On the other hand, these angles are different for the Mn atoms of the
A and B sublattices reflecting their inequivalence in the magnon states.
Figure~\ref{fig_phi}(b) presents for each atom the value of the angle between the $xy$ projection of its
moment and of the moment of the atom lying next below. As seen in figure (a), this is always
the angle between the moments of the Mn and Te atoms. These angles are distinctly 
different for Mn atoms of the A and B sublattices confirming again their inequivalence. 
 
\subsection{Chirality splitting of magnons in altermagnets}
\label{sec_chirality_symmetry}

To study the chirality properties of magnons the approach to the 
symmetry analysis must be fundamentally revised compared to the spin-splitting study
of the electron states discussed above (Sec.~\ref{sec_degeneracy_electron_states}).
Now the analysis is focused not
on the properties of the electron states of the same magnetic configuration
but on the relation between energies of different magnetic configurations.
If we take an arbitrary SSG operation $\{\alpha_S|\alpha_R|\boldsymbol{\tau}\}$
and transform our magnetic system according Eq.~(\ref{eq_SSG_operator})
we obtain the system with the properties directly related to
the properties of the initial system. This  conservation
of the properties 
reflects two factors: first, the homogeneity and anisotropy of the space where the system is placed 
and, second, the invariance under the applied transformation of the form of the interactions taken into account 
in the considered physical model \cite{comment}.
In particular, the equivalence of the systems connected by the SSG transformation reflects
the fact that both space shift and space rotation 
of the system does not change the energy of the system.
To reveal the magnon states having, for symmetry reasons, equal energies
we will act with operations $\{\alpha_S|\alpha_R|\boldsymbol{\tau}\}$ on a selected magnon state 
aiming to determine other magnon states equivalent to it.
Since dealing with the magnon states of the system we are not interested in the copies of the system 
obtained by the displacement of the atomic lattice
or by the rotation of the net sublattice magnetization from the $z$ axis 
the operations we consider are restricted to follows:
the orbital part  $[\alpha_R|\boldsymbol{\tau}]$ of the transformation belongs to the space group
of the crystal and leaves the lattice unchanged whereas the spin transformation $\alpha_S$ keeps the net magnetizations 
of the sublattices collinear to the $z$ axis.  
 
If $[\alpha_R|\boldsymbol{\tau}]$
transforms magnetic sublattices into themselves
the spin part must keep the directions of the sublattice magnetizations unchanged 
$\alpha_S {\bf e}_z$=${\bf e}_z$. Here {\bf e}$_z$ is the unity vector parallel to the
$z$ axis.
Respectively, 
if $[\alpha_R|\boldsymbol{\tau}]$
transforms sublattices into each other
the spin part must reverse the directions of the sublattice magnetizations $\alpha_S {\bf e}_z$=$-$${\bf e}_z$.
This shows
that the set of the operations we need to consider 
forms exactly the SSG of the collinear AFM ground state that is the group
used above (Sec.~\ref{sec_degeneracy_electron_states}) in the discussion of the altermagnetic 
spin-splitting of the electronic states of the
GS collinear configuration. 

This conclusion reveals a direct analogy between the pattern of the spin-splitting 
of the electron states in the $\bf{k}$ space and
the pattern of the chirality splitting of the spin waves in the ${\bf q}$ space.
Indeed, if $\alpha_R$ transforms $\bf{k}$ vector 
of an electron state in $\alpha_R\bf{k}$ it transforms ${\bf q}$ vector of a magnon in  $\alpha_R\bf{q}$. And if 
this operation reverses spin of the electron state it changes also the chirality of the magnon.
Hence if there is spin-splitting (spin-degeneracy) of the electron states at point $\bf{k}$ 
there is also the chirality splitting (chirality degeneracy) of the magnon states for wave vector $\bf{q}$=$\bf{k}$. 
Therefore the reciprocal-space symmetry patterns
of the spin splitting in the ground state electron band structure and chirality splitting of magnons are 
identical. In particular, Fig.~\ref{Fig_stars} reflecting spin splitting of the electron states in the {\bf k} space
is directly applicable to the analysis of the chirality splitting of the magnons in the {\bf q} space.

\subsection{Calculated magnon dispersion}
\label{sec_magnon_calculat}

\begin{figure}[t]
\includegraphics*[width=8.5cm]{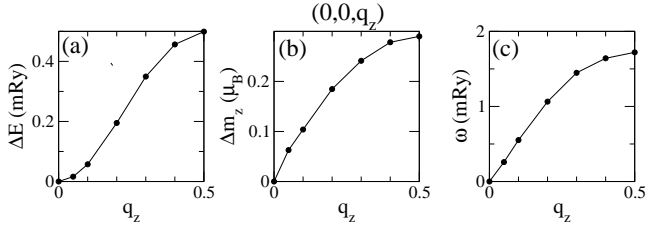}
\caption{Calculation of the spin wave energies in the $\Gamma$A interval of the BZ:
{\bf q}=(0,0,$q_z$), $q_z$$\in$[0,0.5].
(a) $\Delta E$, (b) $\Delta m_z$, (c) spin wave energy $\omega$.
}
\label{Fig_SW_qZ}
\end{figure}

In Fig.~\ref{Fig_SW_qZ} we show the results of the calculation of the spin wave energies in the 
$\Gamma$A interval of the BZ: {\bf q}=(0,0,$q_z$), $q_z$$\in$[0,0.5].
Close to the $\Gamma$ point the energy increase $\Delta E$$\sim$$q_z^2$ whereas  $\Delta m_z$$\sim$$q_z$.
This gives a linear dependence of the spin wave energy $\omega$ on  $q_z$, as expected for the
magnons in AFMs in the region of the $\Gamma$ point. In agreement with symmetry arguments, the magnons of both chiralities are 
degenerate at the wave vectors from this interval. 

\begin{figure}[t]
\includegraphics*[width=8.5cm]{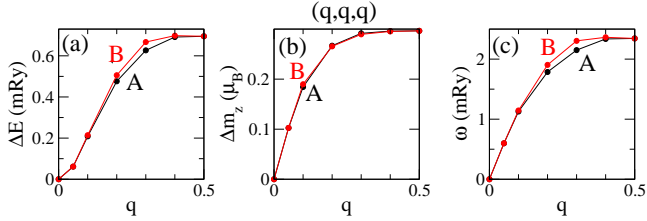}
\caption{Calculation of the spin wave energies in the interval
{\bf q}=($q$,$q$,$q$), $q$$\in$[0,0.5].
(a) $\Delta E$, (b) $\Delta m_z$, (c) spin wave energy $\omega$.
Black curves show the results for magnons of type A, red curves 
for magnons of type B.
}
\label{Fig_SW_qXXX}
\end{figure}

From the symmetry analysis (Sec.~\ref{sec_chirality_symmetry}) it is expected that for the wave vectors 
lying on a less symmetric line we should obtain chirality splitting
of the magnon states. In Fig.~\ref{Fig_SW_qXXX} we show the result of the 
calculations for interval {\bf q}=($q$,$q$,$q$), $q$$\in$[0,0.5].
The calculations confirm the prediction of the symmetry analysis:
numerical difference between data for A and B magnons is 
obtained for all points of the interval with exception for the end points. However,
the difference is small and, for a part of the interval, is not noticeable in the
figure.

\begin{figure}[t]
\includegraphics*[width=8.5cm]{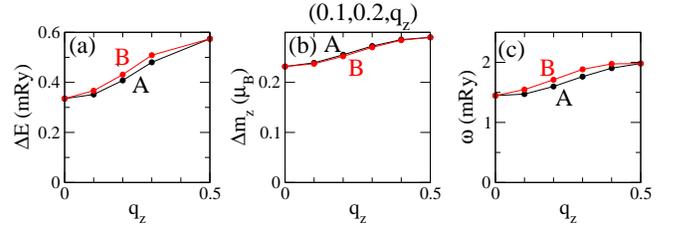}
\caption{Spin wave energies in the interval
{\bf q}=($0.1$,$0.2$,$q_z$), $q_z$$\in$[0,0.5].
(a) $\Delta E$, (b) $\Delta m_z$, (c) spin wave energy $\omega$.
Black curves show the results for magnons of type A, red curves 
for magnons of type B.
}
\label{Fig_SW_q12z}
\end{figure}

In Fig.~\ref{Fig_SW_q12z} we show the result of the 
calculations for an interval {\bf q}=($0.1$,$0.2$,$q_z$), $q_z$$\in$[0,0.5] in the reciprocal
space. This interval contains general points not invariant with respect to any symmetry operation. 
Again the calculations confirm the results of the 
symmetry analysis: at all inner points of the interval there is chirality splitting of the magnon
states. The maximal splitting is $\sim$7\% with respect to the energy of the spin waves at the 
corresponding {\bf q} point.

\textcolor{black}{
\subsection{Instability of the energy of magnon state with respect to the number of magnons}
\label{sec_singular}
\begin{figure}[t]
\includegraphics*[width=4cm]{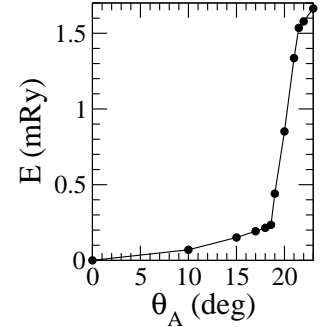}
\caption{Instability of the energy of magnon state with respect to the number of magnons.\label{Fig_jumps}
}
\end{figure}
}

\textcolor{black}{
       While studying the performance of the method we carried out
numerous model calculations. In particular, we investigated the 
consequences of the relative energy shift of the Mn 3d and Te 5p states 
that influences the hybridization between these states. 
Since the nonmagnetic atoms, in our case Te atoms, play crucial role in the 
formation of altermagnetism, the character of the hybridization of the electron states 
of magnetic and nonmagnetic atoms is an important factor influencing the character
of the altermagnetic effects.
As a convenient tool for 
varying the energy positions of the Mn 3d and Te 5p states we employed LDA+$U$
method with various values of the corresponding $U$ parameters.
In these numerical experiments we used not only
parameter $U_{Mn}$ for the Mn 3d states 
but also small $U_{Te}$ parameter for the Te 5p states, the latter treated as a free parameter 
assuming both positive and negative values. 
In these model calculations, we noticed that in some cases the spin wave states behave singular as 
the function of $\Delta m_z$.
}

\textcolor{black}{
In Fig.\ref{Fig_jumps}, we show a typical instability behavior. This result was obtained for the energy of 
the magnon states of type A with
wave vector {\bf q}=(0.1,0.2,0.3) calculated as a function of constrained angle $\theta_A$
with parameters $U_{Mn}$=0.2~Ry, and $U_{Te}$=$-$0.03~Ry.
The increase of $\theta_A$ can be associated with increasing number of magnons.
Around $\theta_A$ of 20$^\circ$ the system experiences sharp transformation between two 
different states. It is straightforward to suppose that in this $\theta_A$ region, the increase
of  $\theta_A$ leads to the change of the relative energy positions of
different groups of electron states in the band structure of the system. 
In the iterations, the consequences of this process 
are enhanced leading to the sharp variation of the energy of 
the magnon state.
}

\textcolor{black}{
A simular unstable behavior we obtained for magnon B as a function of the constrained 
angle $\theta_B$ (not shown). However, the region of instability was shifted to somewhat smaller 
angles. As the result, in the instability region we obtained an enhancement of the chirality splitting effect.
Thus, for $\theta_A$=$\theta_B$=20$^\circ$ the difference in energies of the A and B
magnon states is $\sim$0.6~mRy that is about two orders of value larger than for angles of 15$^\circ$.
Simular instabilty was obtained in calculations for other wave vectors {\bf q}.
The presence of such instabilities cannot be noticed applying the standard procedure of mapping on the 
Heisenberg Hamiltonian. The possibility to study such instabilities that are different for different magnon chirality
channels is a useful feature of the suggested method.
}

\subsection{\textcolor{black}{Relation between electronic band structures of the magnon states 
and chirality properties of these states}}
\label{sec_trans_path}

Both chirality splitting and chirality degeneracy of two magnons manifest certain relation
between the energies of the magnons.
Whether two magnons are equivalent and have equal energies or are inequivalent and have different energies 
must be reflected in the properties of their electron band structures.
In this section we gain an insight into the connection between 
electronic  band structures of the magnon states and chirality properties of these states.
\textcolor{black}{
Our method includes the calculation of the electronic  band structures of the magnons
as a part of the itiration process.
}

The concepts of the generalized periodocity [Eq.~(\ref{eq_sym_constr})] and generalized 
Bloch theorem \cite{Sandratskii1986} will help us to establish
the transformation path from the electronic band structure of the collinear AFM GS to the electronic band
structure of the magnon with a given wave vector {\bf q} and given chirality. 
In this way we reveal the features of the electron energy spectrum connected
with the presence or absence of the chirality splitting.

We begin with the electron band structure of the ground state.
As pointed out in Sec.~\ref{sec_degeneracy_electron_states}, the collinear magnetic ground state can be treated as a  
spiral with  $\theta$=0 and  arbitrary wave vector {\bf q}. 
Different {\bf q} values lead to different relative shifts of the spin-up
and spin-down electron states in the reciprocal space. 
This shift is the consequence of the redefinition of the
wave vector
of the electron state in the case of 
the generalized Bloch theorem
compared to the usual periodicity and usual Bloch theorem\cite{Sandratskii1991}:
depending on spin projection $\sigma$ wave vector of the
electron state changes from $\bf{k}$ to ${\bf k}$$-$$\sigma {\bf q}/2$.
Since in the collinear AFM GS there is no mixing between spin-up and spin-down electron states
the relative shift of these states in the reciprocal space
has no consequences for physically significant characteristics.
However, in the noncollinear magnon state with $\theta$$\neq$0, the wave vector {\bf q} is uniquely defined.
The relative shift by vector {\bf q} of the spin-up and spin-down states is the first step in the transformation
of the band structure that brings the wave vectors of the electron states in agreement with the 
definition according to the generalized Bloch theorem, the only possible way to introduce the wave vector
in the case of noncollinear spiral structures.

The detailed process of the transformation of the shifted GS electronic band structure  to
the electronic band structure of the magnon in a real multiple-band system is complex and
can be obtained only numerically.
However, some important general trends can be distinguished.
One of the consequences of the noncollinearity 
is the hybridization of the spin-up and spin-down electron states.
The strongest hybridization takes place in the regions of the intersection of the spin-up 
and spin-down bands. 
From the symmetry point of view, 
in the collinear ground state the spin-up and spin-down electron states belong to different IRs of the SSG group
and the corresponding bands intersect. In the magnon state the number of the symmetry operations
decreases and spin-up and spin-down states do not any longer belong to different IRs that leads to their
hybridization. The hybridization leads to the repulsion of the bands at the intersection points.

We remark that besides the hybridization repulsion of the bands with opposite spin projections 
in the electron structure of the helical configuration the noncollinearity leads also to 
the transformation of the electron bands with a given spin projection
that has the form of mixing of the ${\bf k}$$-$${\bf q}/2$ and ${\bf k}$$+$${\bf q}/2$
states of the GS electron structure with coefficients 
$\cos^2$$\frac{\theta}{2} $ and $\sin^2$$\frac{\theta}{2} $ (Ref.~\onlinecite{Sandratskii1986}). This effect is quadratic with respect to $\theta$ 
while the spin-mixing effect at the intersection point is linear in $\theta$.

Since the positions of the band intersections are different for different {\bf q}, the process of the hybridization 
is {\bf q} dependent and, therefore, different for different magnons. 
Let us look closer at the transformation of the band structure
for two values of the magnon wave vector {\bf q}: high symmetry vector  ${\bf q}_1=(0,0,0.2)$ 
and general vector ${\bf q}_2=(0.1,0.2,0.2)$.
On the basis of both the symmetry analysis and the calculations discussed in 
Secs.~\ref{sec_chirality_symmetry}, \ref{sec_magnon_calculat}
we know that 
at ${\bf q}_1$ the magnons are chirality-degenerate whereas at ${\bf q}_2$ they are chirality-split.

\begin{figure}[t]
\includegraphics*[width=8cm]{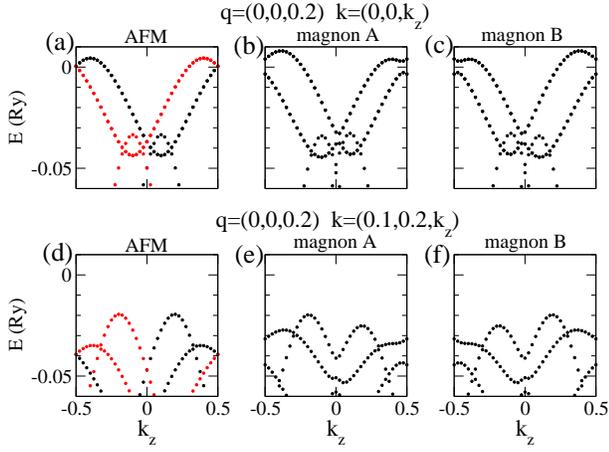}
\caption{The transformation of the electron band structure of the AFM GS to the band structure of magnon states
with ${\bf q}_1=(0,0,0.2)$. Figures (a)-(c) present the bands in the interval [0,0,$k_z$], $k_z\in[-0.5,0.5]$
of the electron BZ. 
The usual AFM band structure in this interval is shown in Fig.~\ref{Fig_bands_AFM}(a). 
Figure (a) gives the AFM band structure 
after the shift of the bands by $\frac{1}{2}\sigma{\bf q}_1 $. Figures (b) and (c) give the electron bands  
for magnon structures of the A and B types. 
Figures (d)-(f) give the band structure in the interval [0.1,0.2,$k_z$], $k_z\in[-0.5,0.5]$.
Corresponding usual AFM band structure is shown in Fig.~\ref{Fig_bands_AFM}(b). Figure (d) gives the AFM band structure 
after the shift of the bands by $\frac{1}{2}\sigma{\bf q}_1$. Figures (e) and (f) give the electron bands 
for magnon structures of the A and B types.
In Figs. (a) and (d), black (red) circles present spin-up (spin-down) states.
}
\label{Fig_bands_SW_q1}
\end{figure}
We begin with ${\bf q}_1$ and consider electron band structure in the
intreval [0,0,$k_z$], $k_z\in[-0.5,0.5]$,
in the {\bf k} space.
On the high symmetry line (0,0,$k_z$) the GS band structure is spin degenerate [Fig.~\ref{Fig_bands_AFM}(a)]. The relative shift 
by  ${\bf q}_1$ results in spin split bands as shown in Fig.~\ref{Fig_bands_SW_q1}(a). 
The change of the sign of $k_z$ combined with the spin reversal
leaves the band structure in Fig.~\ref{Fig_bands_SW_q1}(a) invariant.
The band structures of magnon configurations of both chiralities calculated for wave vector ${\bf q}_1$ and (0,0,$k_z$) line
are presented in Figs.~\ref{Fig_bands_SW_q1}(b),(c). For magnon configuration of type A we used 
angles $\theta_A$=45 and $\theta_B$=0,
for magnon of type B angles $\theta_A$=0 and $\theta_B$=45.
Since in noncollinear structures the electron states are spin-mixed, all states are 
shown in the same color. The band structures of magnons A and B are different and both of them have lost 
the symmetry with respect to the reflection at $k_z$=0. However,
these band structures transform to each other after reflection at $k_z$=0
and, therefore, the integrals over the occupied parts of the spectra for the given {\bf k} interval 
are equal for both magnons.  

Let us continue with the consideration of a low symmetry interval [0.1,0.2,$k_z$], $k_z$$\in$[-0.5,0.5].
The bands have altermagnetic spin splitting in the collinear AFM configuration [Fig.~\ref{Fig_bands_AFM}(b)].
There is the symmetry of the band structure with respect to simultaneous sign change of both $k_z$ and 
spin projection similar to the case shown in Fig.~\ref{Fig_bands_SW_q1}(a) .
After the shift by $\frac{1}{2}\sigma{\bf q}_1$  [Fig.~\ref{Fig_bands_SW_q1}(d)] 
the symmetry between the spin-up and
spin-down bands remains intact. The calculation for the A and B magnon configurations for this {\bf q} gives,
similar to the $(0,0,k_z)$ interval, 
two different band structures [Fig.~\ref{Fig_bands_SW_q1}(e),(f)] which have lost the reflection symmetry. However,
they again transform into each other after reflection at $k_z=0$.
Therefore, also in this case the integrals over occupied parts of the band structures of both magnons are identical. 
These symmetry properties provide an insight
into why the magnons with opposite chiralities
are degenerate at ${\bf q}_1$=(0,0,0.2).

\begin{figure}[t]
\includegraphics*[width=8cm]{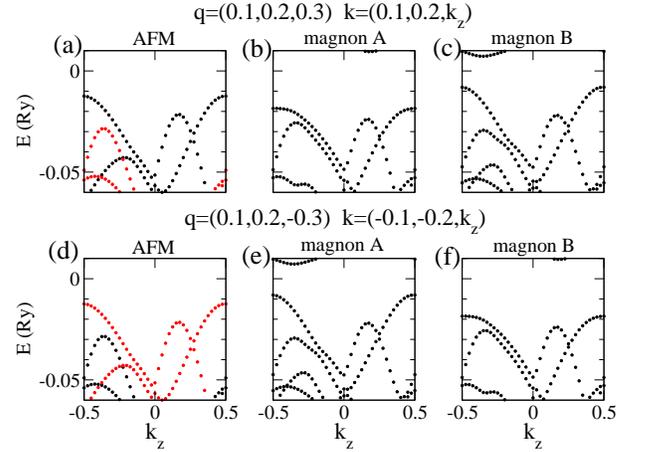}
\caption{The transformation of the electron band structure of the AFM GS to the band structures of magnon states
with wave vectors ${\bf q}_2=(0.1,0.2,0.3)$ and ${\bf q}_3=(0.1,0.2,-0.3)$. 
Figures (a)-(c) give the band structures in the interval  [0.1,0.2,$k_z$], $k_z\in[-0.5,0.5]$ of the electron BZ
for wave vector ${\bf q}_2$.
Figure (a) presents the AFM bands  
after the shift by $\frac{1}{2}\sigma{\bf q}_2$. Figures (b) and (c) give the electron bands  
for magnon structures of the A and B type. 
Figures (d)-(f) show the band structures in the interval [-0.1,-0.2,$k_z$], $k_z\in[-0.5,0.5]$
for wave vector ${\bf q}_3$. 
Figure (d) gives the AFM bands 
after the shift by $\frac{1}{2}\sigma{\bf q}_3 $. Figures (e) and (f) give the electron bands 
for magnon structures of the A and B type.
In Figs. (a) and (d), black (red) circles present spin-up (spin-down) states.
}
\label{Fig_bands_SW_q2}
\end{figure}
Now let us consider the magnons with low symmetry wave vector ${\bf q}_2$=(0.1,0.2,0.3) that is not invariant 
with respect to any space group transformation. 
We take again the $[0.1,0.2,k_z]$ interval of the BZ. 
After $\frac{1}{2}\sigma {\bf q}_2 $ 
shift in the GS band structure [Fig.~\ref{Fig_bands_SW_q2}(a)] we obtain in this interval the spin-up and 
spin-down bands that are essentially different and 
cannot be transformed into each other by a symmetry operation.
Respectively, magnons A and B result in this case in two essentially different
band structures [Figs.~\ref{Fig_bands_SW_q2}(b),(c)]
giving different energy contributions
to the integral band energies of the A and B magnons. The difference of the band structures of the A and B 
magnons with wave vector ${\bf q}_2$   
illustrates the origin of the chirality splitting of the magnons at this wave vector.

On the other hand, for  ${\bf q}_3$=(0.1,0.2,$-$0.3) in  the $[-0.1,-0.2,k_z$] interval
we obtain the band structures [Figs.~\ref{Fig_bands_SW_q2}(d)-(f)] very similar to those
shown in Figs.~\ref{Fig_bands_SW_q2}(a)-(c).
Figure ~\ref{Fig_bands_SW_q2}(d) for the AFM GS is
identical to Fig.~\ref{Fig_bands_SW_q2}(a)  after reversal of the spin projectrions of all electron states.
The band structure of magnon A with wave vector ${\bf q}_3$ [Fig.~\ref{Fig_bands_SW_q2}(e)] is identical
to the band structure of magnon B  with wave vector ${\bf q}_2$ [ Fig.~\ref{Fig_bands_SW_q2}(c)].
The band structure of magnon B with wave vector ${\bf q}_3$ [Fig.~\ref{Fig_bands_SW_q2}(f)] is identical to 
to the band structure of magnon A  with wave vector ${\bf q}_2$ [ Fig.~\ref{Fig_bands_SW_q2}(b)].
These properties of the band structures
illustrate the band-structure basis of the chirality degeneracy of the magnons with different wave vectors
and chirality splitting at given wave vectors.  

The following point is worth emphasizing: As noted above, in the band structures of magnons 
(Figs.~\ref{Fig_bands_SW_q1},\ref{Fig_bands_SW_q2}) the electron states are spin mixed
and, therefore, all bands are presented in the same color. 
However, the contributions of the spin-up and spin-down components of the spinor eigenfunctions are 
different for different states. Some of the electron states can remain almost spin-up or almost spin-down.
On the other hand, in the regions where spin-up and spin-down bands intersect in the GS the hybridization 
leads to the electron states with large contributions of both spin projections.
For magnons, the integrated spin contributions of the occupied electron states do not completely compensate each other 
since each magnon has a nonzero magnetic moment. 
 
\section{Conclusions}
\label{sec_conclusions}
In the paper we suggest the method for direct ab initio calculation of magnons in complex collinear
magnets 
\textcolor{black}{
without
}
the spin-orbit coupling. The method does not include the mapping 
of the electron system on the Heisenberg Hamiltonian of the interacting atomic moments as an 
intermediate step.

Each magnon state is obtained in a separate
DFT based self-consistent calculation performed under two different constraints. One constraint 
governs the magnetization change with respect to the ground state and the other is a symmetry 
constraint establishing the value of the magnon wave vector {\bf q}. 
The performance of the method is demonstrated by 
the application to an altermagnet MnTe. The main feature of the GS of the altermagnets is the spin splitting
of the electron states with a given wave vector {\bf k}. Also the magnons in altermagnets have a special feature:
chirality splitting of magnons with the same wave vector {\bf q} that is the energy difference of the magnons
corresponding to the different magnetic sublattices. We show that despite different nature of the spin-splitting
of the electron states and chirality splitting of the magnon states have identical patterns in the 
corresponding wave vector spaces. All conclusions based on the symmetry arguments are confirmed by the 
results of numerical calculations.

\textcolor{black}{
Our method allows investigation of the electron band structure of individual magnons.
}
We investigate the connection between the chirality properties of the magnons and 
the properties of the electron band structures of the magnons. 
The deep connection between these two different energy characteristics is exposed.

The altermagnetism of MnTe  is the consequence of the presence of the Te atoms. 
\textcolor{black}{
The method allows the study of the self-consistent response of the nonmagnetic 
atoms to the formation of magnon excitations. An adequate
attention is devoted to the analysis of the properties of the Te atoms 
in magnons in MnTe.
}
In particular, we show that the Te atoms remain equivalent in the magnon states though the Mn atoms 
beloning to different sublattices became inequivalent.
\textcolor{black}{
The information about induced magnetic moments of the Te atoms obtained in the symmetry analysis
besides providing important physical information
helps to accelerate the numerical convergence of the magnon states by means of decreasing the number of 
degrees of freedom in the self-consistent calculations.
}

In the paper, different symmetry aspects are analyzed within the framework of the spin space groups.
These aspects include the symmetry constraint of generalized periodicity in the suggested method of the 
magnon calculations, the spin splitting of the electron states in the GS, and chirality splitting of the
magnon states. The SSGs that are the generalization of the usual space groups allow
the integration of these different aspects in one coherent physical picture.

The suggested method accompanied with symmetry arguments on the basis of the spin-space groups
provides a new 
tool 
making possible efficient
direct first principles study of the magnons in complex collinear magnets 
including emerging class of altermagnets.

The spin splitting in some of the AFM crystals was known for many years but only few years ago it became the focus
of intense research efforts within the new research field dubbed altermagnetism.  Possible applications of the
altermagnetic spin splitting are under discussion.
On the other hand, the chiral splitting of the magnon states 
that is a fundamental feature of altermagnets has been noticed very recently. 
Though important from the theoretical point of view,
this splitting seems to be relatively weak. However, also this feature
of altermagnets can become important for future experimental studies and practical applications 
in the field of magnonics. A necessary step on this path is the search for materials with large chirality splitting.
Our method that self-consistently takes into account the crucial contribution of the nonmagnetic atoms
provides a useful tool in this search. Also, the ability of the method to reveal possible chirality-sensitive instabilities 
of the magnon excitations can prove itself useful in this respect.

\textcolor{black}{
An important question is if the method can be applied to the systems where the SOC cannot be neglected. 
The main problem with the SOC is that it reduces the 
6D space with separate spin and orbital variables to the actual 3D space
where any point transformation acts on both types of variables. In the presence of SOC, 
the generalized translations do not commute with the Kohn-Sham Hamiltonian
and the reduction of the problem, for an arbitrary wave vector {\bf q}, to the small crystallographic unit cell 
cannot be performed. This mathematical problem is the reflection
of the physical reality consisting in the fact that the SOC disturbs the perfect 
helical structure of the spin waves by means of the magnetic anisotropy.
However, there is an important class of materials where the method can potentially be applied in the presence of the SOC. These are the materials with uniaxial magnetic anisotropy. If the $z$ axis is an easy axis and the anisotropy in the 
$xy$-plane is negligible, the spin rotation by an arbitrary angle about the anisotropy axis leaves the system invariant, and the generalized translation symmetry remains intact.
In the Ref.~\onlinecite {Sandratskii2017}, the possibility to neglect the in-plane anisotropy  
was used to study the Dzyaloshinskii-Moriya interaction that is one of the consequences of SOC. 
The extension of the direct method suggested in this paper to the case of the magnets with uniaxial anisotropy is an interesting problem for the future studies. Another possibility to deal with the SOC is, first, to perform self-consistent calculation of the magnon states without SOC and then to take into account the SOC within the first-order perturbation theory \cite{Heide2009}.
}
\section{Acknowledgements}
The discussions with Samir Lounis are gratefully acknowledged.
K.C. acknowledges financial support by Czech Science Foundation, grant no. 23-04746S, and project Quantum materials for applications in sustainable technologies (QM4ST), funded as project no.  CZ.02.01.01/00/22\_008/0004572 by P JAK, Ministry of Education,
Youth and Sports of the Czech Republic.
V.M.S. acknowledges financial support by grant PID2022-139230NB-I00 
funded by MCIN/AEI/10.13039/501100011033.


\begin{thebibliography}{99}
\bibitem{comment_Niu}
\textcolor{black}{
We remark that 
the contribution of Ref.~\onlinecite{Niu99} was to show that  Eq.~\ref{eq_omega}, although expected from the 
general concept of spin waves can be derived in the DFT context. 
In this respect, Ref. ~\onlinecite{Niu99}
contributes to the theoretical fundament of our method. 
However, the mathematical apparatus of Berry curvatures used in Ref. ~\onlinecite{Niu99} does not provide
a basis for a practical method for a direct DFT-study of magnons in complex materials.  
To suggest such a method is one of the tasks of this paper.
}
\bibitem{comment}
The formula of the SOC is not invariant with respect to the $\{\alpha_S|\alpha_R|\boldsymbol{\tau}\}$ operations 
with different $\alpha_S$ and $\alpha_R$.
Therefore in the relativistic case the new system obtained by transformation $\{\alpha_S|\alpha_R|\boldsymbol{\tau}\}$ 
is essentially different from the initial one.
\end{thebibliography}
\end{document}